\documentclass[shortauth]{aa}

\usepackage{graphicx}

\usepackage{txfonts,textcomp}

\usepackage[colorlinks=true,
            linkcolor=blue,
            urlcolor=blue,
            citecolor=blue]{hyperref}

\usepackage{orcidlink}

\def\RJ{R_\mathrm{J}}
\def\MJ{M_\mathrm{J}}

\begin{document}

   \title{Oxygen and nitrogen isotopologues on cold COCONUTS-2b observed with MIRI/MRS} 

    \author{H. Kühnle\orcidlink{0009-0001-2738-2489}\inst{1}  \and 
    E. C. Matthews\orcidlink{0000-0003-0593-1560}\inst{2} \and
    P. Mollière\orcidlink{0000-0003-4096-7067}\inst{2} \and
    P. Patapis\orcidlink{0000-0001-8718-3732}\inst{1} \and
    Z. Zhang\orcidlink{0000-0002-3726-4881}\inst{3} \and
    E. Nasedkin\orcidlink{0000-0002-9792-3121}\inst{4} \and
    D. Gasman\orcidlink{0000-0002-1257-7742}\inst{2,5} \and
    N. Whiteford\orcidlink{0000-0001-8818-1544}\inst{6} \and 
    H. S. Wang \orcidlink{0000-0001-8618-3343}\inst{7} \and
    M. Ravet \orcidlink{0009-0000-4898-4713}\inst{8,9,2} \and
    G. Chauvin \orcidlink{0000-0003-4022-8598}\inst{2} \and
    M. Bonnefoy \orcidlink{0000-0001-5579-5339}\inst{9} \and
    D. Barrado\orcidlink{0000-0002-5971-9242}\inst{10} \and 
    A. M. Glauser\orcidlink{0000-0001-9250-1547}\inst{1} \and
    S. P. Quanz\orcidlink{0000-0003-3829-7412}\inst{1,11} }
 
   \institute{
ETH Zürich, Institute for Particle Physics and Astrophysics, Wolfgang-Pauli-Str. 27, 8093 Zürich, Switzerland 
\and
Max-Planck-Institut für Astronomie, Königstuhl 17, D-69117 Heidelberg, Germany
\and
Department of Physics \& Astronomy, University of Rochester, 500 Joseph C. Wilson Blvd. Rochester, NY 14627
\and
Trinity College Dublin, The University of Dublin, College Green, Dublin 2, Ireland 
\and
Institute of Astronomy, KU Leuven, Celestijnenlaan 200D, 3001 Heverlee, Belgium
\and
Department of Astrophysics, American Museum of Natural History, New York, NY 10024, USA
\and
Centre for Star and Planet Formation, Globe Institute, University of Copenhagen, Øster Voldgade 5-7, 1350 København, Denmark
\and
Laboratoire J.-L. Lagrange, Université Côte d'Azur, Observatoire de la Côte d'Azur, CNRS, 06304 Nice, France
\and
Institut de Planétologie et d'Astrophysique de Grenoble (IPAG)/CNRS, Université Grenoble Alpes, 621 avenue Centrale, 38400 Saint-Martin-d'Hères, France
\and
Centro de Astrobiología (CAB), CSIC-INTA, ESAC Campus, Camino Bajo del Castillo s/n, 28692 Villanueva de la Cañada, Madrid, Spain
\and
Department of Earth and Planetary Sciences, ETH Zürich, Sonneggstrasse 5, Zurich 8092, Switzerland
}

    \date{Draft version \today}

  \abstract
   {Linking the composition of gas giant planets to their formation paths has long been a goal in exoplanet science. Especially, cold gas giants with temperatures below $\sim$500K have been out of reach for detailed atmospheric characterization. With JWST, however, we can reach high signal-to-noise (S/N) spectra for such cool worlds and can can measure not only their main trace gas abundances, but even their isotopic content unlocking new possibilities in linking them to their formation paths. In this study, we present the spectrum of one of the coldest planetary-mass companions COCONUTS-2b ($\mathrm{T_{eff}}\approx$480K, separation of $\sim$6400 au from its M dwarf host star) obtained with the Mid-InfraRed Instrument Medium Resolution Spectrometer (MIRI/MRS). We receive a high S/N spectrum of up to 40 at $\sim$ 11.8 µm. Combining the MIRI and archival Gemini/FLAMINGOS-2 data sets, we aim to characterize the chemical composition and physical structure of its frigid atmosphere, setting the stage to uncover insights on the formation of COCONUTS-2b.
   For the first time on a MIRI/MRS data set, we use the full spectral resolution of MIRI/MRS and perform atmospheric retrievals to unlock the search for faint absorption features by rare molecules and isotopologues. The latter are identified using a leave-one-out analysis and Bayes factor comparison.
   We robustly detect three isotopologues, namely  $^{15}$NH$_3$, H$_2^{18}$O and H$_2^{17}$O in the atmosphere of COCONUTS-2b. We find the first clear evidence of oxygen isotopes in water in a cold companion, complementing previous CO isotope detections. We constrain the $^{16}$O/$^{18}$O and $^{16}$O/$^{17}$O isotope ratios to be significantly enriched compared to the solar and ISM values, however the value for $^{18}$O/$^{17}$O is consistent with the ISM value. We find a nitrogen ratio $^{14}$N/$^{15}$N compatible with the value of the ISM and with previous observations of Y brown dwarfs. The derived effective temperatures as well as subsolar C/O and sub- to solar metallicity are in line with previous results.
   This data set demonstrates the capability of MIRI/MRS to characterize such cold planetary-mass companion's atmospheres with respect to their compositional and isotopic content. In the future, the constrained elemental and isotope ratios provide a unique avenue in comparing with the host star's abundances and eventually in tracing formation scenarios. }

   \maketitle

\section{Introduction} 

The atmospheric physics and oxygen, nitrogen, and carbon chemistry of cold worlds beyond our Solar System contextualize the corresponding properties of Jupiter. For a long time, we have not been able to characterize them in detail due to their faint fluxes in the accessible wavelengths. With the advent of the James Webb Space Telescope \citep[JWST, ][]{Gardner2023,Rigby2023} we now obtain their spectra in the mid-infrared (MIR) in unique spectral resolving power and signal-to-noise (S/N) providing valuable information about the composition. This allows us to unravel not just features of the major absorbing species, but even their faint isotopologues. 

Understanding the formation of gas giants similar to Jupiter requires profound knowledge of the properties of such objects \citep[e.g.][]{Burrows1997,Knierim2025}. Recent discoveries  of cold gas giant companions with direct imaging by the Mid-InfraRed Instrument on board JWST \citep[JWST/MIRI,][]{Wright2015,Wright2023} have populated a region of the parameter space previously inaccessible, such as, Eps Ind Ab, with an estimated effective temperature of $\sim$275 K \citep{Matthews2024}, the similarly cold companion 14 Her c \citep{Bardalez2025} or a sub-jovian planet in the young TWA 7 debris disk \citep{Lagrange2025}. More detailed characterizations of these cold worlds are also possible as shown by the the clear detection of ammonia in the atmosphere of GJ504b \citep{Malin2025}.

All these discoveries provide promising insights into the formation, evolution and dynamics of Jupiter-like planets, but in practice, extracting high S/N spectra from direct observations suffer from the contrast to their host star. Thus, objects orbiting far away from their host stars are promising and easier targets to characterize their atmospheric composition and structure. Further, free-floating brown dwarfs are also laboratories to study the atmospheric properties of cold objects: their compositions, temperatures, and the governing physical processes that are thought to be comparable with cold gas giants \citep[e.g.][]{Burrows1997, Coulter2022}. 

While the spectra of brown dwarfs at the L/T transition have been studied intensively \citep[e.g.][]{Miles2020,Vos2023,Biller2024,Molliere2025,Nasedkin2025}, the T/Y transition is still rather unexplored. With JWST we are now able to probe this cold end of the brown dwarf population \citep{Barrado2023,Beiler2024a,Beiler2024b,Matthews2025,Kuhnle2025,Vasist2025,Luber2026}. The chemistry of brown dwarfs at the T/Y transition with temperatures of $\sim$500 K \citep{Cushing2011} is dominated by water, ammonia and methane and show strong absorption lines in the mid-infrared (MIR). To date, only one T-dwarf spectrum measured with MIRI/MRS has been published. The first T dwarf observed with MIRI/MRS showed the unexpected presence of HCN and C$_2$H$_2$ \citep{Matthews2025} and has a similar temperature compared to COCONUTS-2b.

A newly emerging avenue to probe formation paths of cold brown dwarf and gas giant atmospheres is enabled by studying isotopologues \citep{Barrado2023,Gandhi2023,Kuhnle2025,Ruffio2026}. The unprecedented resolving power of the medium resolution spectrometer (MIRI/MRS) \citep{Wells2015, Wright2015, Argyriou2023} allows to do so in the MIR. Isotopic ratios have been proposed as a formation tracer for cold gas giants and brown dwarfs, as it may allow us to link the atmospheric composition to the disk composition during formation \citep{Molliere2019a,ZhangYapeng2021a,ZhangYapeng2021b,Barrado2023,Nomura2023}. Oxygen isotopes are abundant in the Inter Stellar Medium (ISM) and the Solar System and have been measured extensively \citep[e.g.][]{Wilson1999,McKeegan2011}. The local region in which our Solar System is located has been found to be enriched in $^{18}$O in comparison to the average ISM from ejecta from supernovae of type II \citep{Young2011}. \citet{Gandhi2023} detected the oxygen isotopologues of CO in the planetary mass object VHS1256b, and \citet{Ruffio2026} also detected the same in the HR8799 planets c, d and e.

Typically, these compositional results are derived using atmospheric retrievals, which have been widely used as a tool for the characterization of exoplanets and brown dwarfs \citep[e.g.,][]{Madhusudhan2009,Molliere2020,Barrado2023,Vos2023,Faherty2024,Kothari2024,Nasedkin2024b,Rowland2024,Matthews2025,Vasist2025,Nasedkin2025}. 

Here, we present a retrieval analysis of the MIRI/MRS spectrum of the widely separated companion COCONUTS-2b using the publicly available retrieval code \texttt{petitRADTRANS} \citep{Molliere2019,Blain2024,Nasedkin2024a}.

COCONUTS-2b is the one of the furthest planetary-mass objects from its host star and thus is a prime target to study the atmosphere of a wide-separation planet \citep{Zhang2025}. COCONUTS-2b was discovered as one of the first WISE targets by \citet{Kirkpatrick2011} as a field brown dwarf. \citet{Zhang2021} showed using the GAIA early data release 3 and as part of the COol Companions ON Ultrawide OrbiTS survey that it is bound to the M3V dwarf L34-26 (also known as COCONUTS-2A) with a projected separation of $\sim$6471 au or 594'', and is at a distance of 10.888 $\pm$ 0.002 pc \citep{Bailer-Jones2021}. \citet{Zhang2025} presented a Gemini/FLAMINGOS-2 on the Gemini-South telescope spectrum ranging from 1-2.5 µm with an average resolving power of R$\sim$900. They used thermal evolution models, COCONUTS-2b's estimated age of 150-800~Myr and bolometric luminosity ($\rm log(L_{bol}/L_{\odot}) = -6.18 \pm 0.25$ dex) to constrain its bulk parameters, including an effective temperature of T$_{\rm eff} = 483^{+44}_{-53}$K, a surface gravity of $\rm log(g)=4.19^{+0.18}_{-0.13}$ dex, a radius of $\rm R=1.11^{+0.03}_{-0.04}\RJ$ and a mass of $\rm M=8\pm2 \MJ$ \citep{Zhang2025}. They also compared the Gemini/FLAMINGOS-2 spectrum of COCONUTS-2b to sixteen grids of atmospheric models to estimate its sub- to near solar metallicity and C/O ratio. The latter classifies COCONUTS-2b to be a planetary-mass object. Based on its J, H and K band flux it was classified as a T9.5 dwarf and is thus spectroscopically right at the transition between T and Y dwarfs. In addition, it shows signs of disequilibrium chemistry and clouds \citep{Zhang2025}. Recently, \citet{Kiman2025} found that the COCONUTS-2 system is likely a member of the Corona of Ursa Major moving group and thus has an age of about 414$\pm$23Myr \citep{Kiman2025}. Using this age and an updated bolometric luminosity, they find a more precise mass for the companion of 7.5 $\pm$0.4$\MJ$. The composition and bulk properties of the host star have not been widely studied. \citet{Kiman2025} constrain the host stars mass to be 0.4$^{+0.01}_{-0.02} \rm M_{\odot}$ and \citet{Hojjatpanah2019} find the metallicity to be consistent with the solar value (with $Z=0.00\pm$0.08). The lack of information on the host stars composition makes comparisons challenging, especially with respect to isotopic ratios. However, precise measurements of the ISM and the sun as well as recent discoveries in other brown dwarfs and gas giants provide crucial references.

Not only the atmospheric composition, but also the large separation of COCONUTS-2b from its host star raises the question of the formation of the companion. Possible scenarios include: 1) the companion has formed like a star through gravitational collapse \citep{Boss1997}, 2) it has formed inside a disk of the host star through gas accretion \citep{Pollack1996} and migrated outwards to its position observed today or 3) it has evolved completely independent of the host star and was captured during a fly-by \citep{Zhang2021,Marocco2024}. Free-floating planets might be much more frequent than expected in a star-forming region \citep{Miret-Roig2022}, lending credence to the possibility of such a capture event. However, the COCONUTS system belongs to a looser moving group making a fly-by capture rarer. Further, the bulk parameter estimates are consistent with the age of the star, compatible with a gravitational collapse or a disk formation scenario; they are in line with both hot- and cold-start models and thus were not able to rule out either scenario \citep{Zhang2025}. 

In this paper, we present the MIRI/MRS spectrum reaching from 4.9 to 18 µm with a resolving power of up to R$\sim$3750 of COCONUTS-2b. This data set provides the highest resolving power spectrum in this wavelength range observed of COCONUTS-2b to date allowing us to characterize the companions chemical composition and atmospheric structure using full resolution atmospheric retrievals. We introduce the observation and data reduction in Section \ref{data}. The retrieval setups are presented in Section \ref{method}. We present our results of the analysis in Section \ref{results} and discuss in \ref{discussion}. Section \ref{conclusion} concludes this work by summarizing and providing an outlook for future work. 

\section{Observation and data reduction} \label{data} 

\begin{figure*}[h]
    \centering
    \includegraphics[width=\linewidth]{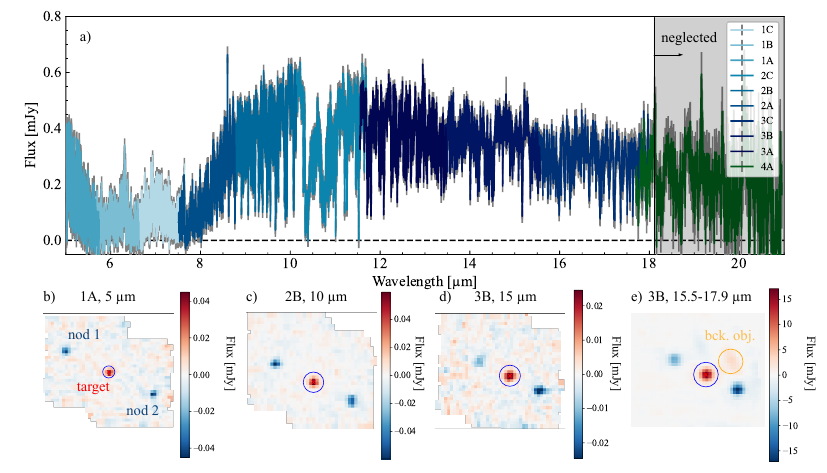}
    \caption{Observation and cube detector images of channel 1A to 4A, where channel 4A gets neglected from further analysis (see text). The red cross corresponds to the position of the target and the blue circle to the aperture from which the flux has been extracted from. The two negative spots on the detector correspond to two nods originating from the background subtraction. The yellow circle shows the position of a background object.}
    \label{fig:det}
\end{figure*}

COCONUTS-2b was observed on 05 September 2024 as part of the GO proposal number 6463 (PI: P. Patapis) using the MIRI/MRS instrument onboard of JWST. The observations consisted of 200 groups per integration, one integration per exposure and one exposure. In total, 3.7 hours of telescope time were allocated for the observations and all three channels (SHORT, MEDIUM, LONG) were used to obtain a complete spectrum. To improve pointing accuracy, target acquisition was performed on the object. The science observation was performed in a negative 4-pt dither pattern in the FASTR1 readout. The observed spectrum with the corresponding detector images at 5, 10, 15 µm and the summed channel 3B from 15.5 to 17.9 µm are presented in Fig. \ref{fig:det}.

The data reduction has been carried out using the JWST pipeline \citep{Bushouse2025} version 1.18.0, CRDS context files jwst\_1364.pmap (version 12.1.6). Stage 1 was reduced from the \_uncal files using the standard steps (including the ramp fitting, dark current subtraction, saturation, linearity and jump corrections). Here, data gets processed from raw Digital Numbers (DN) to \_rate files in DN/s. Stage 2 includes the flat field, stray light, fringe and photometric correction. We added the `clean showers' option to the stray light correction, which improved significantly remaining striping artifacts due to cosmic rays that were not mitigated by the jump correction in stage 1. We use the nod subtraction method on the detector images to remove residual background. Here we subtract the dither pairs with the largest intersection to minimize PSF overlap from each other and replace the original image with the subtracted one. Together with the `clean showers' step mentioned above this results in an exceptionally well subtracted background for this point source as seen in the cube images in Fig. \ref{fig:det}. We included the standard fringe correction, but not the residual fringe correction. We tested for the latter, and the corrections were mainly within the pipeline error bars. Due to the small effect on the spectrum and as it may as well remove actual spectral features \citep{Gasman2023}, we neglect it in this reduction. The final stage 3 creates a cube from the two-dimensional detector image using the `drizzle' algorithm. To extract the spectrum we use the position in the cube determined by the extract\_1d.ifu\_autocen() function for all channels. As the signal to noise dropped significantly in channel 4 we decided to use only up until band 3C for this analysis. As COCONUTS-2b is a quasi-free-floating object and the host star is well outside the field of view, there is no residual light that needs to be subtracted. Thus, we set a circular aperture around the source with an radius of 1.5 $\times$ FWHM of the Point Spread Function (PSF). This number was chosen small enough to not include too much background systematics and large enough to limit sampling artifacts \citep{Law2023}. The cubes show an additional source on the top right compared to the point source marked in a yellow circle. This likely is a background galaxy that is bright in the MIR wavelengths, but not in the near infrared. 

In the analysis presented, we include the archival Gemini/FLAMINGOS-2 observation presented in \citet{Zhang2025} to complete the spectrum in the near-infrared (NIR). We exclude regions in the spectrum with low SNR, namely areas between 1.12-1.18 µm, 1.35-1.45 µm, 1.66-1.95 µm and 2.3-4.95 µm in the retrieval analysis in line with \citet{Zhang2025}. More information on the data reduction can be found in \citet{Zhang2025}. Thus, we obtain a medium resolution spectrum from 1 to 18 µm with gaps between 2.3 and 4.9 µm and in the NIR.

\section{Methodology} \label{method} 
In this study, we use the atmospheric retrieval framework petitRADTRANS \citep[version 3.1.2]{Molliere2019,Blain2024,Nasedkin2024a} to characterize the atmosphere of COCONUTS-2b in terms of chemistry and physical properties based on the MIRI and Gemini data. 

In our initial step, we present two retrievals on a binned spectrum of spectral resolution up to $\frac{\lambda}{\Delta \lambda}=1000$ using c-k opacity. The two retrievals vary by their Pressure-Temperature (PT) parameterization: First, we retrieve the temperature using ten pressure nodes, equally spaced in logarithmic pressures. The temperature at each node is parameterized as a free parameter, a factor $a$ of the temperature at the node below \citep[see][]{Barrado2023,Kuhnle2025}. Note, we allow inversions by choosing the prior range for each temperature node to be between [0.2, 1.2]. We interpolate between the nodes using a spline function. We do not regularize the profile, so no penalization on the change in slope is implemented. This will be denoted as `free' in the following. Secondly, we use another PT parameterization incorporating knowledge from self-consistent models to the retrievals as shown in \citet{Zhang2023}. The retrieved values correspond to a deep temperature at the highest pressure node and ten nodes with values for dlogT/dlogP. We derive the nodes from the Sonora Elf Owl grid models \citep{Mukherjee2024} with $\rm T_{eff}$ in [275, 800] K, logg in [3.25, 4.5] dex, [M/H] in [-1,+1] dex, CO in [0.5, 1, 2.5] times solar and logarithmic $\rm K_{zz}$ in [2,4,7,8,9] according to the procedure presented in \citet{Zhang2023} for pressure levels of 10$^3$ to 10$^{-3}$ and has been extended for lower pressures of 10$^{-6}$ bar in \citet{ZJ2025b}. We present the priors in Table \ref{tab:priors} and the retrieval using this parametrization will be denoted as `constrained' in the following. We vary the PT parameterization here to further analyze the effect of the change in PT structure on retrieval runs at full resolution. 

In both retrievals, we include the treatment of error inflation to capture uncertainties in the model and a potential underestimation of the errors on the data. We retrieve for a value $b$, which is added to the measured error: $\sigma^2 = \sigma_{\mathrm{data}}^2 + 10^b$ as presented in \citet{Line2015} and has been used in various previous setups \citep[e.g.][]{Barrado2023, Matthews2025}. We individually retrieve a value of $b$ for each of the three channels in the MIRI data set (1, 2 and 3) and the Gemini data values separately, adding four more parameters to the retrieval. To test whether this treatment is needed, we performed retrievals without this inflation and could see a significant decrease in the posterior width. When neglecting the error inflation, the retrievals become overconfident due to the small error bars. Therefore, by incorporating an error inflation we are able to mitigate this effect to some extent. 

The chosen priors for the presented retrievals are shown in Table \ref{tab:priors}. For all retrievals, we choose priors derived from evolutionary models on the surface gravity and the radius based on Table 4 in \citep{Zhang2025}, corresponding to a 3$\sigma$ range of retrieved values from an extensive grid-model comparison.

\begin{table}[h]
    \centering
    \caption{List of parameters and priors used in the presented retrievals.}
    \begin{tabular}{l|l}
        \textbf{Low res retrievals} & \textbf{Prior} \\ \hline
        Radius [$\RJ$] & $\mathcal{U}(1.05,1.19)$\\
        log g [dex] & $\mathcal{U}(3.7,4.4)$ \\ 
        log CH$_4$ [\#] & $\mathcal{U}(-10,-0.5)$\\
        log H$_2$O [\#] & $\mathcal{U}(-10,-0.5)$\\
        log CO$_2$ [\#] &$\mathcal{U}(-10,-0.5)$\\
        log CO  [\#] & $\mathcal{U}(-10,-0.5)$\\
        log H$_2$S  [\#] & $\mathcal{U}(-10,-0.5)$\\
        log NH$_3$ [\#] & $\mathcal{U}(-10,-0.5)$\\
        log PH$_3$ [\#] & $\mathcal{U}(-10,-0.5)$\\
        log K  [\#] & $\mathcal{U}(-10,-0.5)$\\
        log Na (a) [\#] & $\mathcal{U}(-10,-0.5)$\\ 
         \hline
        \textbf{Free PT} & \textbf{Prior} \\ \hline
        
        T$_{\rm deep}$ [K] & $\mathcal{U}(100,9000)$\\
        $a$ $\cdot$ T$_{i+1}$ for i $\in \{1,...,9\}$ [\#] & $\mathcal{U}(0.2,1.2)$\\
        
        \hline
        \textbf{Constrained PT} & \textbf{Prior}  \\ \hline
        
        T$_{\rm deep}$ [K] & $\mathcal{U}(100,9000)$\\
        dlogT/dlogP$_1$ [\#] & $\mathcal{N}(0.24,0.015)$\\
        dlogT/dlogP$_2$ [\#] & $\mathcal{N}(0.26,0.015)$\\
        dlogT/dlogP$_3$ [\#] & $\mathcal{N}(0.28,0.015)$\\
        dlogT/dlogP$_4$ [\#] & $\mathcal{N}(0.3,0.01)$\\
        dlogT/dlogP$_5$ [\#] & $\mathcal{N}(0.18,0.03)$\\
        dlogT/dlogP$_6$ [\#] & $\mathcal{N}(0.16,0.03)$\\
        dlogT/dlogP$_7$ [\#] & $\mathcal{N}(0.19,0.05)$\\
        dlogT/dlogP$_8$ [\#] & $\mathcal{U}(-0.05,0.1)$\\
        dlogT/dlogP$_9$ [\#] & $\mathcal{U}(-0.05,0.1)$\\
        dlogT/dlogP$_{10}$ [\#] & $\mathcal{U}(-0.05,0.1)$\\
    \end{tabular} \\
    Note: (a) This abundance is only used for the retrievals in c-k opacity, and not when using the full resolution.
    \label{tab:priors}
\end{table}

In a second step, we use the retrieved outputs of the free and constrained retrieval to enable running the retrievals on the full resolution of the data sets applying line-by-line (lbl) opacities. This allows us to test for trace molecules and their isotopologues using the full resolving power of MIRI/MRS while still including the Gemini data set. As running retrievals on full resolution is much more computationally expensive compared to running them on a binned spectrum, we need to make some simplifications. As we want to probe for trace molecules, we are interested in the chemical composition. Thus, we only retrieve for molecular abundances and fix the PT structure, the radius and surface gravity values, according to the outputs of the lower resolution retrievals. In a statistical sense, we are using the data twice and thus, we expect to reach more confident results in the second compared to the first set of retrievals. As the MIRI MRS instrument resolution varies, we calculate the forward model at a resolution of $\lambda/\Delta \lambda = 15’000$ (about 5 times the instrument resolution), which then gets convolved to the mean instrument resolution per channel \citep{Argyriou2023} and binned to the same spacing as the data.

We perform a leave-one-out (LOO) analysis on the latter, running one retrieval including all molecules we test for, the so-called `base' retrieval, and eight retrievals where we exclude each one molecule. This enables us to do a Bayes factor comparison and thus determine whether the corresponding molecule is significantly improving the fit or not. We run two base retrievals for both the constrained and the free PT parameterization and we again retrieve for the error inflation parameter to fine-tune their values using the full resolution. For all the retrievals presented in the LOO analysis we fix the error inflation values to the ones retrieved in the base retrieval. Tests showed that the values remained within one sigma for the $b$ values in the retrievals with one molecule removed compared to the base, and thus fixing the error inflation parameter saved valuable computation time.

For the retrievals we include the following opacities in c-k resolution (R$\sim$1000) on the binned spectrum: H$_2$O \citep{Rothman2010}, CH$_4$ \citep{Hargreaves2020}, CO \citep{Rothman2010}, CO$_2$ \citep{Yurchenko2020}, NH$_3$ \citep{Coles2019}, H$_2$S \citep{Azzam2016}, PH$_3$ \citep{Sousa-Silva2015}, K \citep{Molliere2019,Allard2016} and Na \citep{Allard2016}. We include as well collisional induced absorption line lists accounting for H$_2$-H$_2$ \citep{Borysow2001,Borysow2002} and H$_2$-He \citep{Borysow1988,Borysow1989a,Borysow1989b} collisions. 
We neglect scattering in the presented retrievals as we deal with a clear atmosphere as we discuss in Section \ref{atmosphere} and \ref{discussion_clouds}. We compare the retrieved abundances to a pre-computed grid of chemical equilibrium calculations implemented in petitRADTRANS, easyChem \citep{Molliere2017,Lei2024} in Section \ref{chem}.  

For the full resolution retrievals, we use the lbl line lists of R$\sim$10$^6$,  convolved and binned down to the corresponding resolving power of the instrument. We use the following opacities: H$_2$O \citep{Polyansky2018}, CH$_4$ \citep{Hargreaves2020}, NH$_3$ \citep{Rothman2013}, H$_2$S \citep{Azzam2016}, PH$_3$ \citep{Sousa-Silva2015}, K \citep{Molliere2019,Allard2016}, $^{15}$NH$_3$ \citep{Rothman2013}, H$_2^{17}$O \citep{Rothman2013}, H$_2^{18}$O \citep{Rothman2013}, $^{13}$CH$_4$ \citep{Gordon2022}, CH$_3$D \citep{Rothman2013}, HDO \citep{Rothman2013}, C$_2$H$_2$ \citep{Rothman2013} and HCN \citep{Harris2006}. We include the same collisional-induced absorption lines and Rayleigh scattering as for the lower resolution retrievals. We retrieve for the abundances of  H$_2$O, H$_2^{18}$O, H$_2^{17}$O, HDO, CH$_4$, $^{13}$CH$_4$, CH$_3$D, CO$_2$, CO, NH$_3$, $^{15}$NH$_3$, H$_2$S, PH$_3$, C$_2$H$_2$, HCN and K. We chose the list of isotopologues based on the three most abundant trace gases: H$_2$O, CH$_4$ and NH$_3$. In addition, we test for C$_2$H$_2$ and HCN as they have been detected unexpectedly in a T dwarf of similar temperature \citep{Matthews2025}. We neglect Na in the full resolution retrievals as it was unconstrained for the lower resolution retrieval and only has very low opacity in the wavelength range we probe with MIRI/MRS as shown in Fig. \ref{fig:app_opacities} in the appendix.

\begin{figure*}[h]
    \centering
    \includegraphics[width=\linewidth]{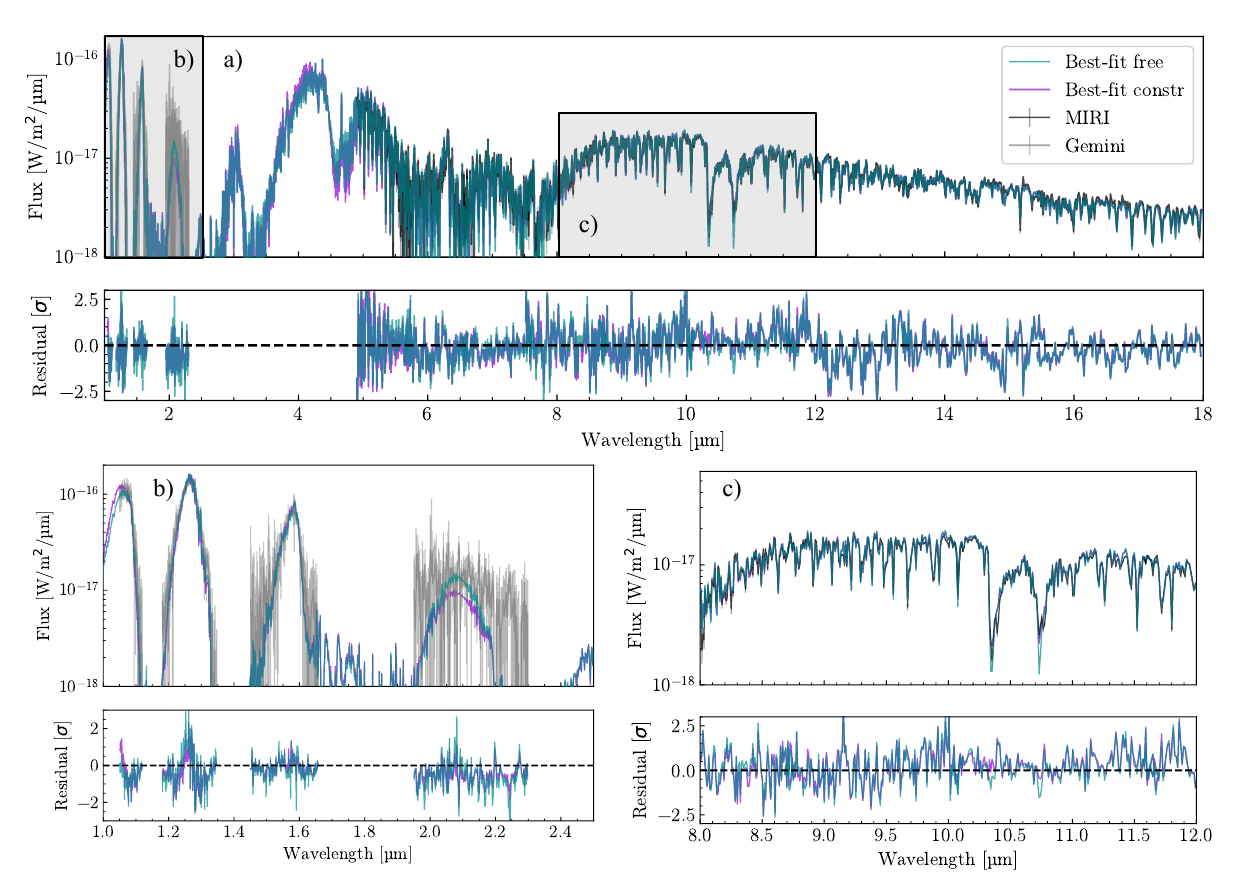}
    \caption{Best-fit spectra of the free and constrained atmospheric retrievals in green and violet respectively compared to the data in black. Panel a) shows the full observation. In panel b) we show a subset at the Gemini wavelengths from 1 to 2.5 µm as indicated by the black box and in c) we show another subset for the strong NH$_3$ feature between 8 and 12 µm. The spectra are shown in lower resolution $\frac{\lambda}{\Delta \lambda}=1000$. }
    \label{fig:fit}
\end{figure*}

To sample the prior space we use the nested sampling algorithm \texttt{MultiNest} \citep{Skilling2004,Feroz2007} with the corresponding python implementation given by \texttt{pyMultinest} \citep{Buchner2014}. We run all retrievals using 1000 live points with a sampling efficiency of 0.05. All full resolution retrievals are run in non-constant efficiency mode to probe the evidence precisely enabling a valid Bayes factor comparison, whereas the lower resolution retrievals are run in constant efficiency mode, as we are here not interested in the exact values of the evidence.  

To decide whether a molecule is detected or not we use the Bayes factor \citep[e.g.][]{Benneke2013, Thorngren2025}. We apply the method as presented in \citet{Konrad2024} computed depending on the evidences of the two best-fits of the base $\mathcal{Z}_b$ and the one excluding the molecule in question $\mathcal{Z}_m$: $\rm log(\mathcal{B}) = \frac{ln(\mathcal{Z}_b)-ln(\mathcal{Z}_m)}{ln(10)}$. Large values in $log(\mathcal{B})$ correspond to the molecule significantly improving the fit and thus being detected. Values for $log(\mathcal{B})$ below 0.5 show no significant improvement and negative values even favor the model that excludes the molecule \citep{Thorngren2025}. 

In addition, we calculate the BPICS value \citep{Ando2011,Thorngren2025} as a different metric for the leave-one-out analysis. Here, smaller values indicate a preference of the model. 
The BPICS is a simplified version of the Bayesian Predictive Information Criterion \citep[BPIC,][]{Ando2007} and based on \citet{Thorngren2025} computed like: $\rm BPICS = -2E_p [ln(\mathcal{L})] + 2N_p$, where $\rm N_p$ is the number of parameter used for the model (ranging between 16 and 15 for the base and the leave-one-out retrieval respectively) and $\rm E_p[ln(\mathcal{L})]$ the median value of the natural logarithm of the likelihood $\mathcal{L}$ (assuming a Gaussian distribution of the latter).

\section{Results} \label{results} 

In this section we present the results of our retrieval analysis on the combined MIRI/MRS and Gemini/FLAMINGOS-2 data. We will start with discussing the bulk parameters and PT structure, then the chemical composition and how it compares to the chemical equilibrium. Further, we will discuss the full resolution retrieval with emphasizes on isotopologues.

\subsection{Bulk parameters} \label{bulk}
In Figure \ref{fig:fit} we show the best-fit spectra of the retrievals using the free and the constrained PT profiles compared to the MIRI/MRS data in black and the Gemini/FLAMINGOS-2 data in grey. Panel a) shows the spectrum from 1 to 18 µm (with gaps between 1.12-1.18 µm, 1.35-1.45 µm, 1.66-1.95 µm and 2.3-4.95 µm to exclude low S/N areas in the spectrum), panel b) a subset on the Gemini data and c) a subset on the ammonia feature in the MIRI/MRS data set. We see both fits explain the general spectrum well. The free retrieval finds a better agreement of the 2.2 µm peak, which however is a part of the spectrum with larger uncertainty on the data. The constrained retrieval fits the depth of the ammonia feature in the region between 8 and 12 µm slightly better. The largest differences between the two best-fit models are between 2.5 and 5 µm due to the lack of data. Especially, the retrieved abundance of CO is slightly different as the CO feature is deeper for the constrained case compared to the free case. Larger differences can be seen at 5-6 µm, probably originating from the CO abundance, not being matched precisely. The rest of the spectrum is explained well by both models and the residuals are on the order of 1$\sigma$ from the data, when including the error inflation.

\begin{figure*}[h]
    \centering
    \includegraphics[width=\linewidth]{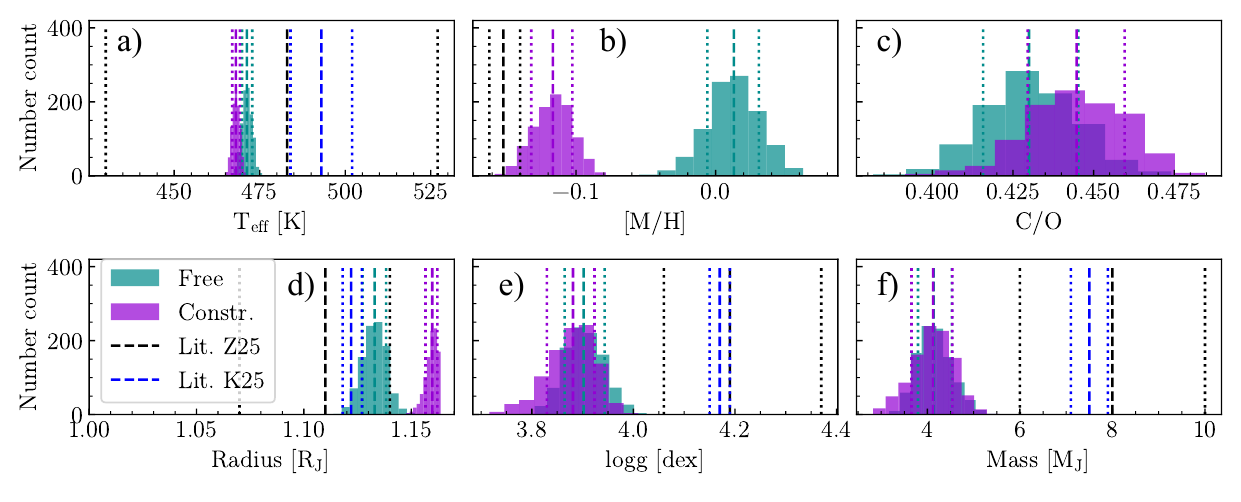}
    \caption{Bulk parameter for the free and the constrained retrievals. In panel a) we show the effective temperature estimate in b) the retrieved  metallicity, in c) the C/O ratio accounting for oxygen sequestration, in d) the radius, in e) the surface gravity and in f) the resulting mass based on the radius and gravity for the free and constrained retrieval in green and violet respectively. For panels a),b) and d)-f) we show in black dashed lines the mean values and in dotted lines the one sigma estimate obtained by \citet{Zhang2025} (Z25) and in blue by \citet{Kiman2025} (K25).}
    \label{fig:bulk}
\end{figure*}

The obtained bulk parameters for the free and constrained retrievals are presented in Fig. \ref{fig:bulk}. We obtain an effective temperature estimate using the built-in function in petitRADTRANS (\texttt{compute\_effective\_temperature}) integrating model spectra from 1 to 18 µm, which were sampled from the posterior. The effective temperature estimates fall well inside the one sigma range presented from the grid models \citep{Zhang2025} shown in black dashed and dotted lines (mean and error, respectively). The free retrieval results in an effective temperature of 471.3$^{+1.6}_{-1.4}$ K compared to the value of 483$^{+44}_{-53}$ K presented in \citet{Zhang2025}. This is compatible within three sigma with the value returned from the constrained retrieval of 468.1$^{+1.1}_{-1.1}$ K. \citet{Kiman2025} present a slightly warmer estimate with a value of 493$^{+9}_{-9}$ K. In general, the values found in this work and from \citet{Kiman2025} fall in the estimated range by \citet{Zhang2025}.

The metallicity is calculated based on the retrieved abundances using the built-in function \texttt{volume\_mixing\_ratios2metallicity} in petitRADTRANS. We find a solar to sub-solar metallicity of 0.01$^{+0.02}_{-0.02}$ dex and -0.12$^{+0.01}_{-0.02}$ dex in the free and constrained case, respectively, which is in line with the findings of \citet{Zhang2025} finding a solar to subsolar metallicity. Their best-fitting model gave a metallicity of -0.152$^{+0.012}_{-0.010}$ dex. The difference in the metallicity is likely connected to changes in the PT structure as it will be discussed in Section \ref{atmosphere}. The free retrieval is consistent with the host stars solar metallicity of 0.00$\pm$0.08 \citet{Hojjatpanah2019}.

The C/O ratio is calculated from all carbon bearing species (namely CH$_4$, CO$_2$ and CO, based on the estimates of the low resolution retrievals) divided by the oxygen bearing molecules (H$_2$O, CO and CO$_2$) accounting for the multiplicity of the respective atoms. 
For colder brown dwarfs, \citet{Calamari2024} determined an amount of $17.8^{+1.7}_{-2.3}\%$ of the oxygen budget being locked up in low lying condensing clouds using stoichiometric and mass balance calculations. Thus, we multiply our obtained C/O ratio by a factor of 1.22 to account for this oxygen sink according to the procedure presented in \citet{Rowland2024}. The free and constrained retrievals constrain a slightly sub-solar C/O ratio with a value of 0.43$^{+0.02}_{-0.01}$ and 0.45$^{+0.02}_{-0.02}$, respectively. \citet{Zhang2025} finds a tendency to favor a subsolar C/O ratio, which is in line with our findings.

We retrieve the surface gravity and the radius and can calculate a mass from these parameters. The radius in the free retrieval is constrained as 1.133$^{+0.005}_{-0.006}$ $\RJ$ and in the constrained case as 1.160$^{+0.002}_{-0.003}$ $\RJ$. Both radii are slightly enlarged compared to the value of 1.07$^{+0.03}_{-0.04}$ $\RJ$ found by \citet{Zhang2025}. In the retrieval, we added an evolutionary prior to the radius from 1.04 to 1.19 $\RJ$. The logarithmic surface gravity of the free retrieval is estimated as 3.90$^{+0.04}_{-0.04}$ dex and in the constrained case 3.88$^{+0.04}_{-0.05}$ dex. Again, we apply prior boundaries for the surface gravity between 3.7 and 4.4 dex and based on the three sigma adopted parameter in \citet{Zhang2025}. Both of the gravity values are slightly smaller compared to previous values (4.19$^{+0.18}_{-0.13}$ in \citet{Zhang2025} and 4.17$\pm$0.02 in \citet{Kiman2025}). The mass estimate is calculated from the radius and gravity values, and as our gravity estimates are smaller, we receive a smaller estimate of 4.1$^{+0.4}_{-0.3}$ and 4.1$^{+0.4}_{-0.5}$ $\MJ$ for the free and the constrained retrieval, respectively, compared to the 8$\pm$2 $\MJ$ \citep{Zhang2025} and 7.5$\pm$0.4 $\MJ$ \citet{Kiman2025}. However, this mass estimate is obtained purely through the radiative transfer model, and thus may be influenced due to degeneracies with other parameters. Specifically, it has been shown that surface gravity is difficult to estimate from atmospheric models alone \citep{Mader2026}. Thus, measuring the mass through evolutionary models including a better age and luminosity estimate similar to what has been done in \citet{Kiman2025}, would result in a more realistic mass measurement. To assess the degeneracies in the retrieved bulk parameters, we compare them in the corner plot in Fig. \ref{fig:app_corner} in the appendix. 

In general, we find a solar to subsolar metallicity, a subsolar C/O ratio compatible with \citet{Zhang2025} and an inflated radius, a smaller gravity and a smaller mass in comparison to \citet{Zhang2025} and \citet{Kiman2025}. 

\begin{table*}[h]
    \begin{center}
    \caption{Summary of the bulk parameters shown in Fig. \ref{fig:bulk}.}
    \begin{tabular}{l|l|l|l|l}
         &  Free & Constrained & Literature Z25 & Literature K25  \\ \hline \hline
        T$_{\rm eff}$ [K] & 471.3$^{+1.6}_{-1.4}$ & 468.1$^{+1.1}_{-1.1}$ & 483$^{+44}_{-53}$ & 493$^{+9}_{-9}$\\ 
        Metallicity [dex] & 0.01$^{+0.02}_{-0.02}$ & -0.12$^{+0.01}_{-0.02}$  & -0.152$^{+0.012}_{-0.01}$ & - \\
        C/O & 0.43$^{+0.02}_{-0.01}$ & 0.45$^{+0.02}_{-0.02}$  & - & -\\
        Radius [$\RJ$] & 1.133$^{+0.005}_{-0.006}$  & 1.160$^{+0.002}_{-0.003}$  & 1.07$^{+0.07}_{-0.04}$ & 1.122$^{+0.005}_{-0.004}$  \\
        log(g) [cm/s$^2$] & 3.90$^{+0.04}_{-0.04}$  & 3.88$^{+0.04}_{-0.05}$  & 4.14$^{+0.18}_{-0.13}$ &  4.17$^{+0.02}_{-0.02}$ \\  
        Mass [$\MJ$] & 4.1$^{+0.4}_{-0.3}$  & 4.1$^{+0.4}_{-0.5}$  & 8$^{+2}_{-2}$ & 7.5$^{+0.4}_{-0.4}$ \\
    \end{tabular}
    \label{tab:iso}
    \end{center}
    Note: Z25 corresponds to \citet{Zhang2025} and K25 to \citet{Kiman2025}.
\end{table*}

\subsection{Atmospheric structure} \label{atmosphere}
\begin{figure}[h]
    \centering
    \includegraphics[width=\linewidth]{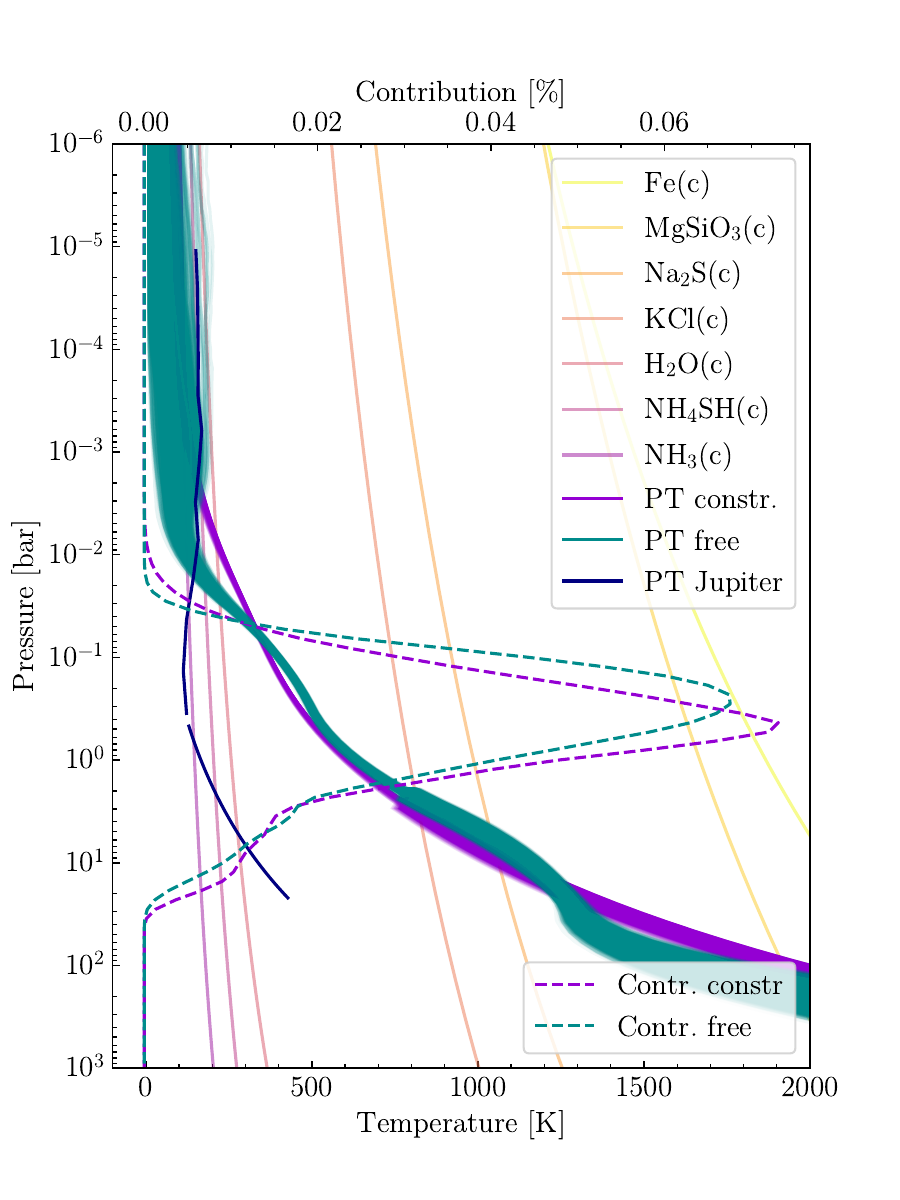}
    \caption{Pressure-temperature profiles of the free and the constrained retrieval in green and violet respectively. In dark blue we show the PT structure of Jupiter. The H$_2$O, NH$_4$SH and NH$_3$ condensation lines were taken from \citep{Lodders2002} and the rest from petitRADTRANS. Thicker lines in deeper atmospheric layers indicate the convective part of the atmosphere calculated based on Eq. \ref{eq:conv}.}
    \label{fig:pt}
\end{figure}

In Fig. \ref{fig:pt} we show the retrieved Pressure-Temperature (PT) structure for the free and constrained retrievals in green and violet respectively. The free retrieval provides a not strictly monotonic PT structure, with areas where the temperature gradient is not constant at around 50 bar, 0.5 bar and 0.01 bar. As we have not use any form of regularization here, this might be due to the spline interpolation between the nodes resulting in such non-monotonies. Further, retrievals tend to overfit the data and such small changes in the PT structure might locally explain the data better. By comparing the free to the constrained PT profile, we see the strong effect of the constraint on the profile. The constrained retrieval results in a monotonically increasing temperature with pressure, as we would expect it for gas giants without any external heat sources. The overall shape is compatible with the free retrieval. Both retrievals become nearly isothermal above about 0.01 bar. In this area, however, the contribution goes close to zero and thus the changes seem not to affect the spectrum. 
In dark blue we add the measured PT profile of Jupiter \citep{Seiff1998} as a reference for an irradiated gas giant. As expected, the profiles are shifted towards higher temperatures by about 420K. Also, the slope of the retrieved temperatures increase faster with pressure in comparison with the latter. We also plot the condensation lines of the most common species, namely NH$_3$, NH$_4$SH, H$_2$O, KCl, Na$_2$S, Fe and silicates. The mean contribution function across all wavelengths showing where in the atmosphere the flux is emitted from is presented in dashed lines for both cases. Most of the flux is emitted from pressures between 10 and 0.1 bar, where no condensation lines are crossed by the PT profiles. Thus, we do not expect to see strong features from condensing clouds in the atmosphere, and indeed we can explain the spectrum well with a clear atmosphere. More about clouds will be discussed in Section \ref{discussion_clouds}. The emission contribution of the free retrieval is shifted slightly to lower pressures compared to the constrained case. At lower pressures spectral features are less pressure broadened. Thus, to obtain the same spectrum in both cases, the abundances of the trace gases need to be increased \citep{Molliere2015}. Therefore, we find a higher overall metallicity for the free compared to the constrained case as presented in Fig. \ref{fig:bulk}.

We approximate the boundary between the convective and the radiative part using the Schwartzschild criterion for convective stability, assuming the ideal gas and hydrostatic equilibrium finding the following expression for the convective part: 
\begin{equation} \label{eq:conv}
   \rm \frac{\partial T}{\partial P} < \frac{RT}{c_p\mu P}
\end{equation}
with T the temperature, P the pressure, R the molecular gas constant (R=8.314 J/mol K), $\mu$ the mean molecular weight (here $\mu \sim$2.33) and c$_p$ the specific heat capacity of a H and He dominated gas (here c$_p\sim$27.5 J/mol K). Thus, we estimate the transition between the convective and radiative part to be at around 2-3 bars. We followed here the approach as presented in \citet{Heng2017}.

\subsection{Chemical composition} \label{chem}
\begin{figure}[h]
    \centering
    \includegraphics[width=\linewidth]{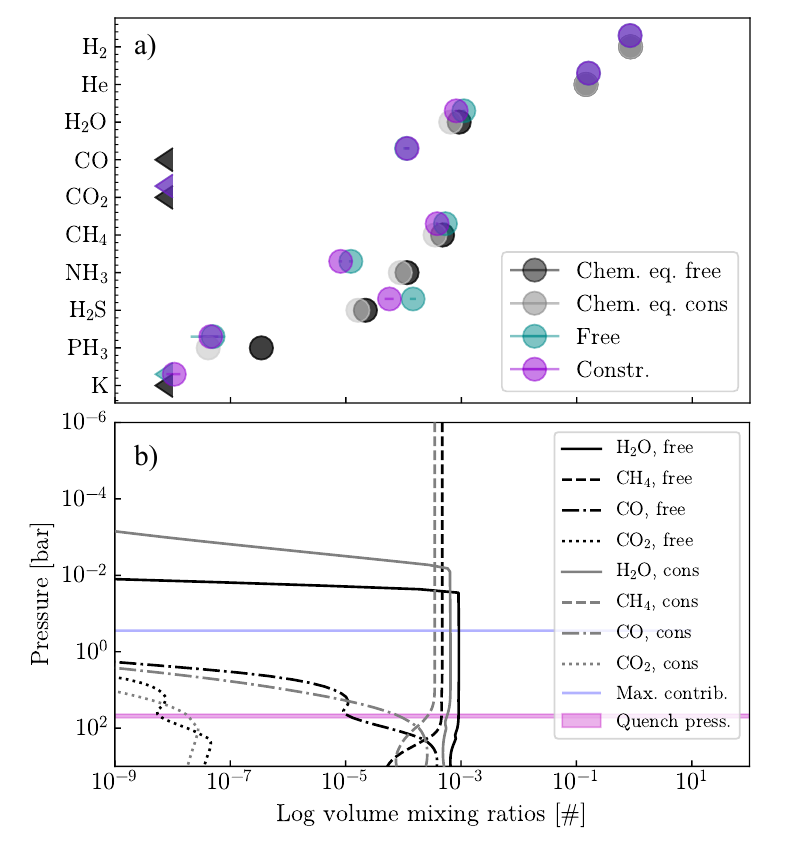}
    \caption{Retrieved chemical abundances with the free and constrained retrievals in spectral resolution $\frac{\lambda}{\Delta \lambda}=1000$ (in green and violet, respectively), compared to the chemical equilibrium predictions for the free and the constrained retrievals (in dark and light grey, respectively). In Panel a) the abundances are taken at 0.28 bar corresponding to the maximum contribution in the free case. In Panel b) the variations with pressure of the abundances involved in the chemical equilibrium, CO; CH$_4$, H$_2$O and CO$_2$ are presented. We additionally indicate the pressure of maximum contribution (0.28 bar) in blue and the quench pressure (43-53 bar) in magenta. }
    \label{fig:chem}
\end{figure}

The strongest absorption features of cold atmospheres below 500K are due to H$_2$O, CH$_4$ and NH$_3$. In our low resolution retrieval analysis we find the following composition presented in Fig. \ref{fig:chem}. In green we show the values for the free retrieval and in violet the constrained values. We report all abundances in logarithmic volume mixing ratios. The exact numbers are given in Tab. \ref{tab:ret_val} in the appendix and we compare the values to the ones from the full resolution retrieval in Fig. \ref{fig:app_abund_fullvslowres}. We compare the retrieved abundances to the chemical equilibrium predictions at 0.28 bar, which corresponds to the height where the maximum emission emerges from in the case of the free PT parameterization. The chemical equilibrium is evaluated based on the retrieved PT structure and the corresponding retrieved metallicity and C/O ratio. In the panel b) in Fig. \ref{fig:chem} we show how the chemical equilibrium changes with pressure for CO, CO$_2$, CH$_4$ and H$_2$O for the free and the constrained case in black and grey respectively. When comparing the retrieved abundances to the equilibrium prediction at the height of maximum contribution (in the free retrieval), we see that they agree well for H$_2$O and CH$_4$. For CO they agree when taking into account quenching at the highlighted quench pressure in magenta. Further, this implies that choosing vertically constant abundances is a valid approximation for the pressure range covered by the emission contribution.

CH$_4$: For CH$_4$ we find a logarithmic abundance of -3.28$^{+0.03}_{-0.03}$ and -3.42$^{+0.02}_{-0.02}$ compared to a slightly lower equilibrium chemistry abundance of -3.33 and -3.46, for the free and constrained case, respectively. The full resolution retrievals obtain a value of -3.247$^{+0.01}_{-0.009}$ and -3.353$^{+0.009}_{-0.009}$ for the free and constrained retrieval respectively, in the same order of magnitude as the low resolution retrieval fits. 

H$_2$O: We find slightly larger values compared to the equilibrium for H$_2$O. Here, we constrain a logarithmic volume mixing ratio of -2.96$^{+0.03}_{-0.03}$ and -3.09$^{+0.02}_{-0.02}$ compared to the chemical equilibrium of -3.04 and -3.18 for the free and constrained case respectively. For the full resolution free and constrained retrievals we find -2.713$^{+0.006}_{-0.007}$ and -2.884$^{+0.006}_{-0.006}$ respectively, compatible with the lower resolution results.

NH$_3$: Larger discrepancies are found for NH$_3$, where we find values of -4.91$^{+0.03}_{-0.03}$ and -5.09$^{+0.02}_{-0.02}$ in the free and constrained case, which are significantly lower compared to the chemical equilibrium -3.94 and -4.06. The full resolution retrievals constrain values for the free and constrained retrievals of -4.925$^{+0.005}_{-0.005}$ and -5.064$^{+0.005}_{-0.005}$ respectively, so similar to the lower resolution fits.

CO and CO$_2$: Further, we find constraints on CO and CO$_2$, which are -3.95$^{+0.05}_{-0.05}$ and -3.94$^{+0.04}_{-0.05}$ for CO and -9.5$^{+0.9}_{-1.0}$ and -9.6$^{+0.9}_{-0.9}$ for CO$_2$ in the free and constrained case respectively. For CO the full resolution retrievals obtain values for the free and constrained cases -4.09$^{+0.03}_{-0.03}$ and -4.06$^{+0.03}_{-0.03}$ and for CO$_2$, -8.4$^{+0.3}_{-0.6}$ and -8.2$^{+0.2}_{-0.3}$ respectively. In chemical equilibrium, abundances of CO and CO$_2$ are negligible at the observed pressure, as temperatures are too low to chemically produce CO and CO$_2$ and rather CH$_4$ is favored. The fact that we constrain these two molecules with values much larger compared to the chemical equilibrium hints that other processes, such as vertical mixing must be present to create the observed disequilibrium chemistry. The implications of this will further be discussed in Section \ref{discussion_dischem}. From the leave-one out comparison we find that CO$_2$ is preferred based on a Bayes factor comparison by 1.07, however due to the only small effect of the molecule on the spectrum as shown in Fig. \ref{fig:isotop_h2s} in the appendix, we conclude a tentative detection. However, another larger band is visible in the 4.2 µm region visible in the NIRSpec data \citep{Kiman2025}, where it could indeed be detected more easily.

From the retrieved values we can calculate the amount of vertical mixing in the atmosphere parametrized by the K$_{zz}$ value. We use the formulas for the chemical timescale where CO and CH$_4$ are in chemical equilibrium and mixing timescales $\rm t_{mix} = H^2/K_{zz}$ with H the scale height based on \citet{Zahnle2014} and rearranged as presented in \citet{Nasedkin2024b}. We obtain a logarithmic K$_{zz}$ of 3.3 cm$^2$/s for both cases at a quench pressure of 43 and 53 bar for the free and the constrained case, respectively.

K and Na: We constrain the alkali species K, but not Na. This might have to do with the very small wavelength range where Na would be visible, which is close to 1 µm and thus at the edge of the Gemini data set and in a noisy area of the spectrum (see \ref{fig:app_opacities}). Both species would not be expected in chemical equilibrium with logarithmic volume mixing ratios much smaller than -10 as they would only be stable in hotter atmospheres. We do not include any rainout of the alkali species and assume constant profiles throughout the atmosphere. As the absorption features of K are only visible between 1-2 µm (see Fig. \ref{fig:app_opacities} in the appendix), this only influences the fit of the Gemini data set. For the free and the constrained case we find an abundance of K of -10.0$^{+0.7}_{-0.7}$ and -8.0$^{+0.1}_{-0.1}$. In the full resolution case we constrain values for K of -10.2$^{+0.8}_{-0.7}$ and -8.1 $^{+0.1}_{-0.1}$ for the free and constrained case respectively.

H$_2$S: For H$_2$S we find a higher logarithmic volume mixing ratio of -3.84$^{+0.05}_{-0.05}$ in the free case and a smaller value of -4.25$^{+0.08}_{-0.08}$ in the constrained case. In the free case the chemical equilibrium prediction is -4.66, and for the constrained case -4.79, which is smaller compared to our retrieved values. This is in line with what we find from the full resolution retrievals, with values of -3.83$^{+0.03}_{-0.03}$ and -4.19$^{+0.04}_{-0.04}$ for the free and constrained case, respectively. 
Including the molecule H$_2$S results in a clear preference according to the logarithmic (base 10) Bayes factor of 61.46 and a clear influence on the flux at around 1.6 µm is visible as shown in Fig. \ref{fig:isotop_h2s} in the appendix. 
The abundance of H$_2$S is slightly larger compared to the chemical equilibrium prediction and varies dependent on the PT choice. As some of the lines that inform the fit are located in the FLAMINGOS-2 data set, which is noisier, this could explain the differences. Also in the NIR, potentially unaccounted for clouds could vary the retrieved abundance when included in the fit. More on clouds will be discussed in \ref{discussion_clouds}. 

PH$_3$: Further, we constrain PH$_3$ in the free case with a value of -7.3$^{+0.2}_{-0.3}$ and in the constrained case a value of and -7.3$^{+0.1}_{-0.2}$. The full resolution retrievals find -7.10$^{+0.06}_{-0.07}$ and -7.47$^{+0.09}_{-0.10}$ for the free and constrained case, being compatible with the lower resolution results. In the chemical equilibrium we would expect a value of -6.46 and -7.36 for the free and constrained case. In equilibrium it drops significantly below pressures of about $\sim$0.3 bar. We find that the full resolution retrieval including PH$_3$ is favored with a logarithmic Bayes factor of 7.03. The value we retrieve is of the order of magnitude we would expect from chemical equilibrium. However, we do not see a clear absorption feature at 10.1 µm as shown in Fig. \ref{fig:isotop_h2s} in the appendix. At this wavelength we would expect to see the largest difference between the model including and the one excluding PH$_3$. Thus, we conclude we only tentatively detect this molecule. The opacity of PH$_3$ might contribute to the continuum and could potentially be influenced by other unaccounted effects, such as clouds. As the presence of this molecule in brown dwarfs is debated, we further discuss this in Section \ref{discussion_dischem}.

C$_2$H$_2$ and HCN: As presented in Table \ref{tab:mols}, we obtain a Bayes factor of 0.19 for C$_2$H$_2$ improving the fit very slightly compared to the base. However, as presented in Fig. \ref{fig:isotop_13ch4}, we do not detect clear absorption features. Therefore, we do not detect this molecule. The logarithmic volume mixing ratios we find show a larger tail towards lower values with logarithmic abundances of -9.1$^{+0.3}_{-0.6}$ and -8.9$^{+0.2}_{-0.5}$ for the free and constrained case. 
HCN was not constrained in the full resolution retrievals with resulting values of -10.1$^{+0.8}_{-0.7}$ and -10.1$^{+0.8}_{-0.7}$ for the free and constrained case, respectively. Using the Bayes factor comparison in the leave-one-out analysis, we find a value of -0.72 and thus conclude that HCN does not significantly improve the fit when included, and we can state a non-detection of HCN. This is confirmed by visual inspection of the fits presented in Fig. \ref{fig:isotop_hcn} in the appendix. We will discuss the absence of the hydrocarbons C$_2$H$_2$ and HCN in Section \ref{discussion_w0458}.

\begin{table*}[h]
    \centering
    \caption{Evidence and BPICS comparison (based on \citet{Thorngren2025}) of the full resolution retrievals of bulk chemical species.}
    \begin{tabular}{l||l|l||l|l||l}
        Retrieval &  $\rm ln(\mathcal{Z})$ & $\rm log_{10}(\mathcal{B})$ & BPICS & $\Delta$ BPICS & Result \\ \hline \hline
        Base & 429108.841±0.009 & 0 & -858294.38 & 0 & \\ \hline
        - H$_2$S & 428967.32±0.01 & 61.46 & -858002.53 & -291.85 & detected \\ \hline
        - PH$_3$ & 429092.65±0.01 & 7.03  & -858257.41 & -36.98  & tent. det.\\
        - CO$_2$ & 429106.37±0.04 & 1.07  & -858287.11 & -7.28   & tent. det.\\ \hline 
        - C$_2$H$_2$ & 429108.40±0.01 & 0.19 & -858291.32 & -3.06 &  undetected \\ 
        - HCN & 429110.499±0.008 & -0.72 & -858296.63 & 2.25 & undetected\\
    \end{tabular}
    \label{tab:mols}
\end{table*}

\subsection{Isotopologues} \label{iso}
To search for isotopologues in the MIRI/MRS spectrum, we perform retrievals on the full resolution of up to R$\sim$3750. We test for isotopologues of the most absorbing species in the infrared (ammonia, methane and water). We included H$_2^{18}$O, H$_2^{17}$O, $^{15}$NH$_3$, $^{13}$CH$_4$, HDO and CH$_3$D and compare the retrievals to the base fit in a leave-one-out analysis. 

\begin{figure*}[h]
    \centering
    \includegraphics[width=\linewidth]{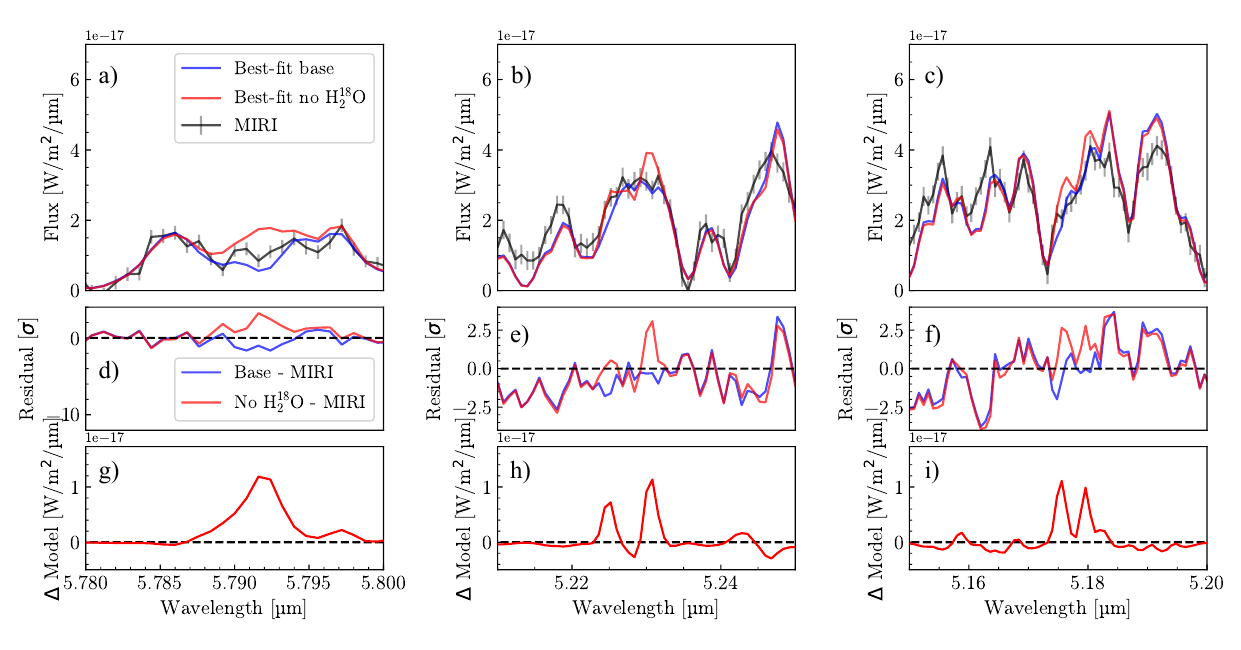}
    \caption{Best-fit retrievals on full MIRI/MRS resolution data with and without including H$_2^{18}$O in blue and red, respectively, compared to the data in black In panel a)-c) we see the three different parts in the spectrum where the difference in the best-fits is visible. The residuals between the data and the corresponding best fits are shown in panels d)-f) and the differences in the fits in plots g)-i). Here, we present the wavelength ranges with fourth, fifth and sixth largest differences in the models as for the first three, the models do not explain the data overall as well. They are presented in the appendix in Fig. \ref{fig:isotop_h218o_all}.}
    \label{fig:isotop_h218o}
\end{figure*}

In Fig. \ref{fig:isotop_h218o}, \ref{fig:isotop_h217o} and \ref{fig:isotop_15nh3}, as well as the corresponding ones in the appendix we show the best-fit spectra for the base retrieval with the constrained PT parameterization without the molecule in question in red and including it in blue compared to the data in black. We also plot the difference between the data and the respective best-fits in the middle plots as well as the difference between the two best-fits in the lowest plots. For the first time in a cold brown dwarf using MIRI/MRS we see absorption features of the water isotopologue H$_2^{18}$O and H$_2^{17}$O in the 5 to 7 µm area as seen in Fig.  \ref{fig:isotop_h218o} and \ref{fig:isotop_h217o}. In addition, including $^{15}$NH$_3$ helps to explain several features in the 9 to 10 µm wavelength range, as presented in Fig. \ref{fig:isotop_15nh3}. More wavelength areas are covered in the appendix.

\begin{figure*}[h]
    \centering
    \includegraphics[width=\linewidth]{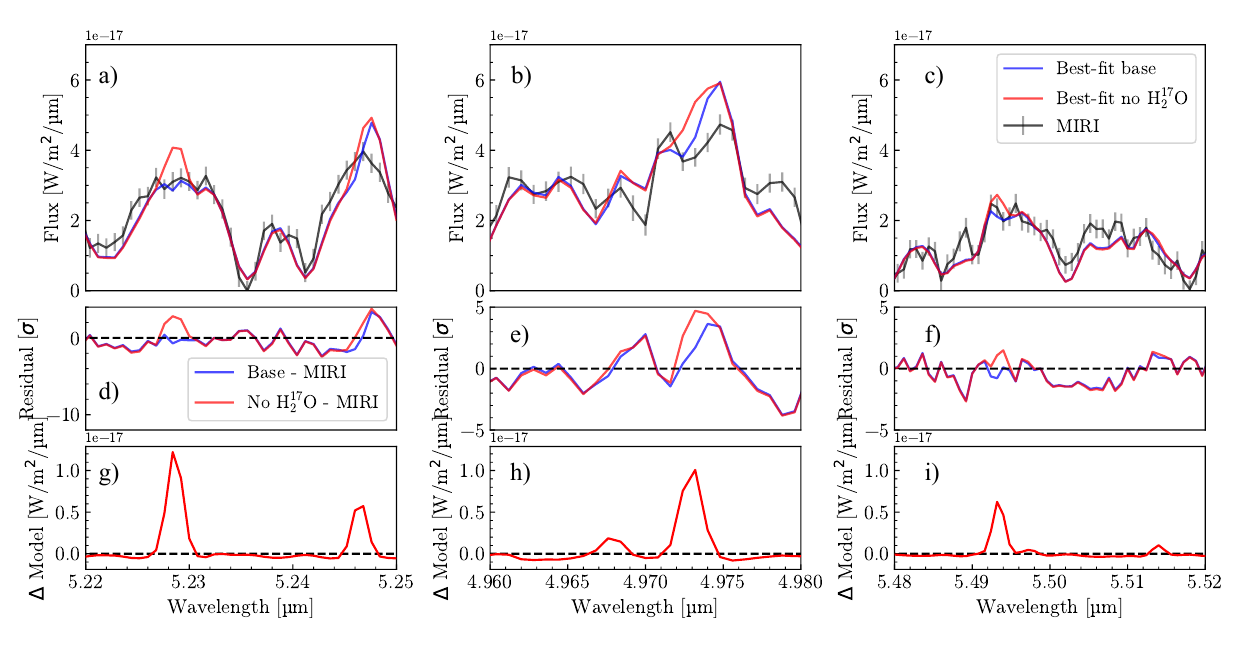}
    \caption{Best-fit retrievals on full MIRI/MRS resolution data with and without including H$_2^{17}$O in blue and red, respectively, compared to the data in black with the same panel structure as in \ref{fig:isotop_h218o}. We present the wavelength ranges with the largest, second and sixth largest differences in the models and more wavelength areas in the appendix in Fig. \ref{fig:isotop_h217o_all}.}
    \label{fig:isotop_h217o}
\end{figure*}
\begin{figure*}[h]
    \centering
    \includegraphics[width=\linewidth]{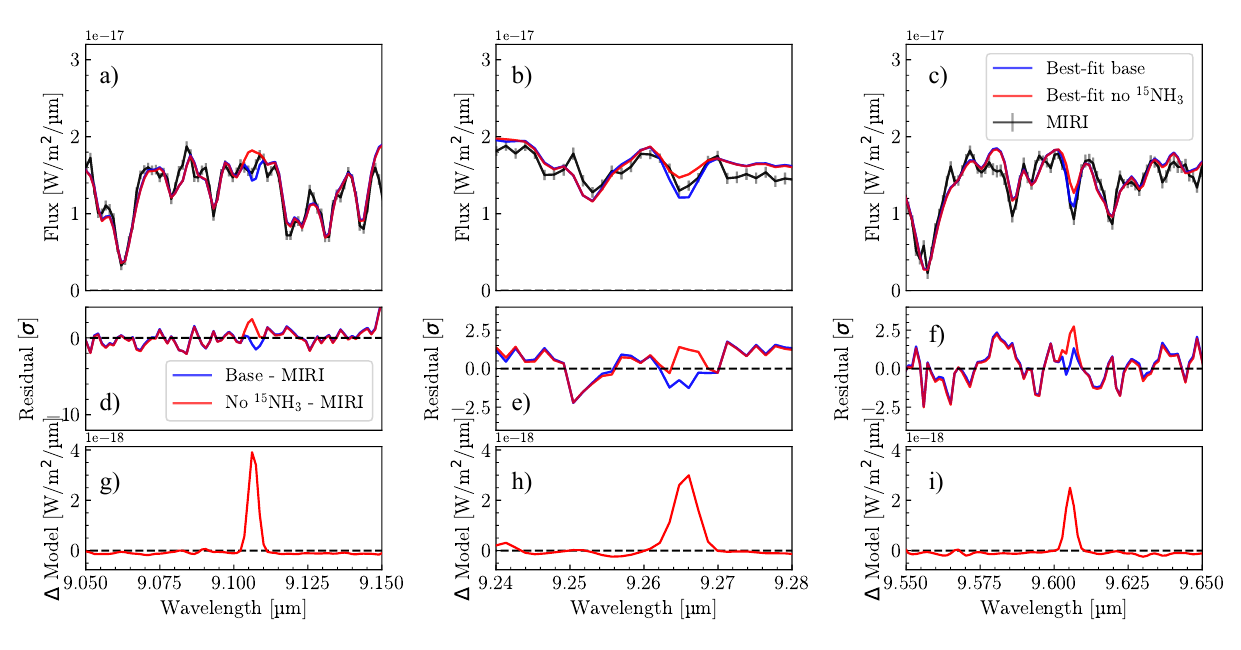}
    \caption{Best-fit retrievals on full MIRI/MRS resolution data with and without including $^{15}$NH$_3$ in blue and red, respectively, compared to the data in black with the same panel structure as in \ref{fig:isotop_h218o}. Here, we present the wavelength ranges with the largest, second and third largest differences in the models and more wavelength areas in the appendix in Fig. \ref{fig:isotop_15nh3_all}.}
    \label{fig:isotop_15nh3}
\end{figure*}

We detect three isotopes confidently with log($\mathcal{B}$) > 20 and a fourth isotope tentatively. In table \ref{tab:iso} we show the logarithmic evidences of the full resolution retrievals comparing the base retrieval with the constrained PT profile to the leave-one-out retrievals. H$_2^{18}$O is favored with a difference of 72.486 compared to the base retrieval, which is the highest value in this comparison. Smaller differences are shown for H$_2^{17}$O with a difference of 25.81 and for $^{15}$NH$_3$ with a difference of 24.68. Thus, these three isotopologues seem to be significantly improving the fit. In comparison, with a Bayes factor of 4.3 the presence of $^{13}$CH$_4$ is only slightly favored in the Bayes analysis. However, as shown in Fig. \ref{fig:isotop_13ch4}, we do not detect clear features of $^{13}$CH$_4$ and therefore state this detection as tentatively. HDO and CH$_3$D are not preferred in the Bayes comparison with negative logarithmic Bayes factors. Also, as shown in Fig. \ref{fig:isotop_hcn} and Fig. \ref{fig:app_abund_fullvslowres} in the appendix, we do not constrain HDO or CH$_3$D, but receive an upper limit, which is further discussed in Section \ref{dh}. As mentioned previously, since we are statistically using the data twice by fixing the PT profile in a second retrieval, the detection significances may be overconfident. Thus, we require to distinguish the absorption features in the spectrum for the detected molecules, which we do for the oxygen and nitrogen isotopologues. As highlighted by \citet{Kipping2025} and \citet{Thorngren2025}, we refrain from discussing the $\sigma$ values in this case.

\begin{table*}[h]
    \centering
    \caption{Evidence and BPICS comparison (based on \citet{Thorngren2025}) of the full resolution retrievals including isotopologues of H$_2$O, CH$_4$ and NH$_3$.}
    \begin{tabular}{l||l|l||l|l||l}
        Retrieval &  $\rm ln(\mathcal{Z})$ & $\rm log_{10}(\mathcal{B})$ & BPICS & $\Delta$ BPICS & Result\\ \hline \hline
        Base &  429108.841±0.009 & 0 & -858294.38 & 0 &\\ \hline
        - H$_2^{18}$O & 428941.935±0.006 & 72.486 & -857954.47 & -339.91 & detected\\
        - H$_2^{17}$O & 429049.41±0.03 & 25.81 & -858169.43 & -124.96 & detected \\
        - $^{15}$NH$_3$ & 429052.01±0.03 & 24.68 & -858176.36 & -118.02 & detected\\ \hline 
        - $^{13}$CH$_4$ & 429098.9±0.1 & 4.3 & -858271.29 & -23.10 & tent. det. \\ \hline
        - HDO & 429109.64±0.03 & -0.35 & -858296.45 & 2.07 & undetected \\
        - CH$_3$D & 429110.22±0.04 & -0.60 & -858296.66 & 2.28 & undetected \\
    \end{tabular}
    \label{tab:iso}
\end{table*}

\begin{table}[h]
    \begin{center}
    \caption{Values for the nitrogen and oxygen ratios obtained with the free and constrained retrievals. }
    \begin{tabular}{l|l|l|l|l}
         &  Free ret. & Constr. ret. & ISM & Solar \\ \hline \hline
        $^{14}$N/$^{15}$N & 337$^{+47}_{-38}$ & 324$^{+46}_{-40}$ & 274$^{+18}_{-18}$ & 458.7$^{+4.2}_{-4.2}$ \\ \hline
        $^{16}$O/$^{18}$O & 247$^{+27}_{-24}$ & 256$^{+29}_{-25}$ & 557$^{+30}_{-30}$ & 529.7$^{+1.7}_{-1.7}$\\
        $^{16}$O/$^{17}$O & 662$^{+98}_{-83}$ & 934$^{+174}_{-139}$ & 2005.2$^{+30}_{-30}$ & 2776$^{+14}_{-14}$ \\
        $^{18}$O/$^{17}$O & 2.7$^{+0.6}_{-0.4}$ & 3.7$^{+0.8}_{-0.7}$ & 3.6$^{+0.2}_{-0.2}$ & 5.5\\ \hline 
    \end{tabular}
    \label{tab:iso_ratios}
    \end{center}
    Notes: References for the ISM values for the oxygen ratios are taken from \citet{Wilson1999}, as well as the solar value for $^{18}$O/$^{17}$O. The ISM value for $^{14}$N/$^{15}$N is from \citet{Ritchey2015}. Solar values for $^{16}$O/$^{18}$O and $^{16}$O/$^{17}$O are from \citet{McKeegan2011} and for $^{14}$N/$^{15}$N from \citet{Marty2011}. Please note that the reported uncertainties might be underestimated due to the fixation of several parameters in the full resolution retrieval.
\end{table}

\begin{figure*}[h]
    \centering
    \includegraphics[width=\linewidth]{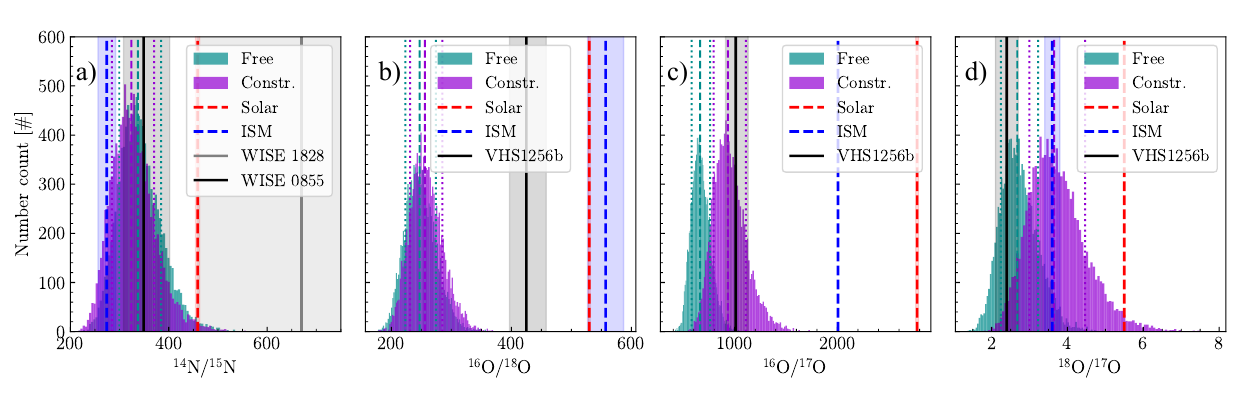}
    \caption{Ratios of a) $^{14}$N/$^{15}$N, b) $^{16}$O/$^{18}$O, c) $^{16}$O/$^{17}$O and d) $^{18}$O/$^{17}$O based on the full resolution retrievals. We compare this to the values for the ISM (Oxygen: \citet{Wilson1999}, $^{14}$N/$^{15}$N: \citet{Ritchey2015}) and solar values ($^{16}$O/$^{17}$O and $^{16}$O/$^{18}$O: \citet{McKeegan2011}, $^{18}$O/$^{17}$O: \citet{Wilson1999} and nitrogen: \citet{Marty2011}) as well to the values of WISE 1828 \citep{Barrado2023}, WISE 0855 \citep{Kuhnle2025} for $^{14}$N/$^{15}$N and VHS1256b for the oxygen ratios \citep{Gandhi2023}.}
    \label{fig:iso_ratio}
\end{figure*}

In Fig. \ref{fig:iso_ratio} we show the isotope ratios $^{14}$N/$^{15}$N, $^{16}$O/$^{17}$O, $^{16}$O/$^{18}$O and $^{18}$O/$^{17}$O derived from the full resolution base retrievals including all molecules using the constrained PT profile. We compare the obtained values to literature values for the sun and the ISM. For nitrogen we additionally compare the retrieved value to the values found from WISE 1828 \citep{Barrado2023} and WISE 0855 \citep{Kuhnle2025} and for the oxygen molecules to the values found in VHS1256b \citep{Gandhi2023} (from C$^{18}$O). In addition, \citet{Ruffio2026} measure C$^{18}$O in three of the four HR8799 gas giants. For $^{16}$O/$^{18}$O they find larger ratios compared to our measurements with significantly larger error bars. The retrieved values are summarized in Tab. \ref{tab:iso_ratios}.

We obtain for $^{14}$N/$^{15}$N values of 337$^{+47}_{-38}$ and 324$^{+46}_{-40}$ in the free and the constrained case, respectively. This is compatible within two sigma with the value found for the ISM of 274$^{+18}_{-18}$ \citep{Ritchey2015} and lower compared to the Solar value of 458.7$^{+4.2}_{-4.2}$ \citep{Marty2011}. We can compare this value also to two Y-dwarfs where these molecules have been found: In WISE1828 a value of 670$^{+390}_{-211}$ \citep{Barrado2023} and 349$^{+53}_{-41}$ \citep{Kuhnle2025} have been constrained (for a clear retrieval). Thus, the values found for COCONUTS-2b are consistent within one and two sigma with the values found in WISE0855 and WISE1828, respectively.

For the ratio $^{16}$O/$^{18}$O we find a value of 247$^{+27}_{-24}$ and 256$^{+29}_{-25}$ for the free and constrained retrieval. These values seem significantly enriched in the heavier isotoplogue when comparing to the ISM value of 557$^{+30}_{-30}$ \citep{Wilson1999} and the solar value of 529.7$^{+1.7}_{-1.7}$ \citep{McKeegan2011}. This ratio was also constrained for the planetary mass companion VHS1256b, with a value of 425$^{+33}_{-28}$ which is significantly larger compared to our findings for COCONUTS-2b.

We constrain the ratio of $^{16}$O/$^{17}$O as 662$^{+98}_{-83}$ and 934$^{+174}_{-139}$ for the free and constrained retrieval compared to the ISM value of 2005.2$^{+30}_{-30}$ \citep{Wilson1999} and the Solar value of 2776$^{+14}_{-14}$ \citep{McKeegan2011}. The value presented in \citet{Gandhi2023} for VHS1256b is compatible with our free case with a value of 1010$^{+120}_{-100}$.

We find the ratio between the two heavier isotopes $^{18}$O/$^{17}$O to be 2.7$^{+0.6}_{-0.4}$ and 3.7$^{+0.8}_{-0.7}$. They are within one to two sigma compatible with the values found for the ISM 3.6$^{+0.2}_{-0.2}$ \citep{Wilson1999}. The values are smaller compared to a solar value of 5.5 \citep{Wilson1999}. Compared to VHS1256b's value of 2.4$^{+0.3}_{-0.3}$ \citep{Gandhi2023}, we find compatible values within one to two sigma for the constrained and free retrieval.

In summary, we find $^{14}$N/$^{15}$N to be compatible with previous observations and the ISM value, $^{16}$O/$^{17}$O and $^{16}$O/$^{18}$O enriched compared to the solar and ISM value and $^{18}$O/$^{17}$O in line with previous observations and the ISM value.

\section{Discussion} \label{discussion} 
\subsection{Full resolution retrievals: Advantages and shortcomings}
 
We present a full resolution retrieval analysis and derive several elemental abundance ratios from it. The reported ratios, however, might be influenced by our assumption to fix the PT profile in the setup. Substantially reducing the number of parameters allows the retrievals to converge in a reasonable amount of time (on the order of days to weeks), and thus is a crucial step to enable this analysis. Generally, the PT structure is determined by the overall flux distribution which is well captured by the lower resolution retrieval. Still, the abundances of the isotopologues might be influenced by small unaccounted changes in the PT structure. To assess the variation in the model we run the same setup with the fixed value for the PT profile from the free and constrained retrievals and compare the derived values. As presented in Fig. \ref{fig:iso_ratio} it has an effect on the ratio we retrieve for $^{16}$O/$^{17}$O, however the ratios of $^{14}$N/$^{15}$N and $^{16}$O/$^{18}$O are compatible with each other within one sigma. 
To investigate how the observed absorption features could be influenced due to changes in the PT structure, we calculated the models with and without the isotopologues for a hotter (positive shift of the interior temperature by 500K), a colder (negative shift by 500K), a steeper (increase in 10\% per node) and a shallower (decrease in 10\% per node) PT profile based on the one retrieved in the constrained retrieval. In Fig. \ref{fig:app_iso_pt} in the appendix we show how the flux varies with respect to the changes. We find that indeed there are differences in the models based on the PT structure, however affecting only the overall shape of the spectrum and not the absorption features of the isotopologues $^{15}$NH$_3$, H$_2^{18}$O and H$_2^{17}$O, as they stay visible for each test case. Thus, we conclude that the PT structure alone could not result in the detected spectral features and thus the assumption of fixing the PT structure is a valid approach. In future studies, more computational power is needed to enable retrieving the PT structure simultaneously with the abundances, which is currently beyond the scope of this setup. 

Further, we fixed the radius and the surface gravity to the values of the lower resolution retrieval. We did this to only retrieve for the abundances simultaneously, as the goal of the leave-one-out analysis was to test for the importance of each molecule. A change in radius only leads to a difference in the absolute flux, and thus is not sensitive to the higher spectral resolution.

The amount of change in the abundances between the low and the full spectral resolution retrievals is assessed in Fig. \ref{fig:app_abund_fullvslowres} in the appendix. Larger differences are seen for H$_2$O, CO and CO$_2$. We retrieve more H$_2$O, less CO and a more contained upper estimate for CO$_2$ when going from low to the full resolution. It is visible that the full resolution retrievals result in narrower posteriors compared to the low resolution results. On the one hand, by fixing several parameter in the full resolution retrievals to values based on the low resolution retrievals, we statistically use the data twice in the setup. Thus, we expect narrower posteriors in the full resolution retrievals compared to the low resolution ones. On the other hand, providing more information from the data in the full resolution case, we receive a more precise estimate of the abundance.

\begin{figure} 
    \centering
    \includegraphics[width=\linewidth]{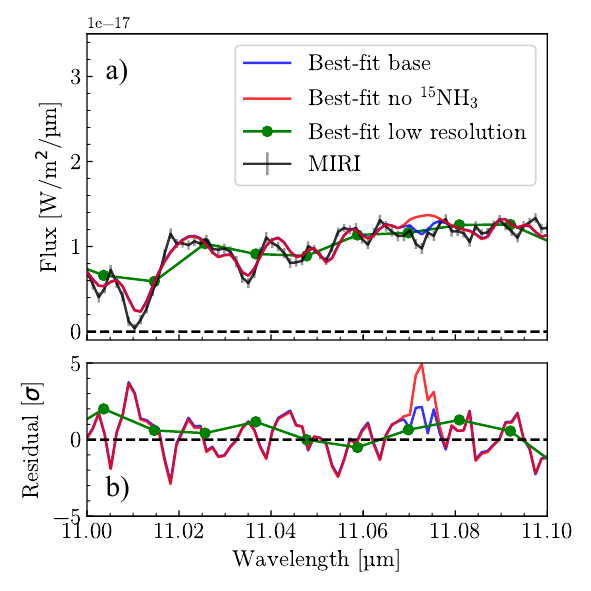}
    \caption{Comparison of low and full resolution retrievals compared to the data. Panel a): The best-fit spectra for the low resolution in green and the full resolution retrievals, with or without $^{15}$NH$_3$ in blue and red, respectively. In panel b) we show the corresponding residuals. Only with the improvement of resolution we are able to detect the isotopologue. }
    \label{fig:fullvslowres}
\end{figure}

In this analysis, we demonstrate that full resolution retrievals are possible using the medium resolution of the MIRI/MRS. Similarly, \citet{Hood2024} and \citet{Ruffio2026} performed medium resolution retrievals on JWST/NIRSpec data. The advantage of running retrievals in full resolution is clearly to detect and measure the abundances of molecules and isotopologues with faint signatures, such as the ones here presented. In Fig. \ref{fig:fullvslowres} we compare the two best fits using the constrained PT profile in the area of 11.0 - 11.1 µm demonstrating the amount of detail gained in the full resolution fits. For this data set, the small spectral signature of $^{15}$NH$_3$ located at 11.075 µm would have not been detectable using the lower resolution retrievals. 
Analyzing the full resolution allows us to study new molecular abundances in current and future medium-resolution observations with JWST. Especially, targets as cold as COCONUTS-2b with high SNR spectra in the MIR will profit from the higher resolution retrievals on MIRI/MRS data. This will help us study compositional differences between host stars and wide separation companions and potentially lead to observational constraints on the formation scenarios of the latter by measuring various elemental abundance ratios in addition to C/O \citep{Oberg2011}. 

\subsection{Small error estimates}
Previous versions of the JWST pipeline underestimated the uncertainties for MIRI/MRS data \citep[e.g.][]{Barrado2023,Kuhnle2025} and the pipeline provided a too small estimate of the error bars. This has now been improved, and the pipeline version used here (v1.18.0) now propagates the errors properly. Small error bars on the data result in difficulties for the retrieval to fit the data as it tends to overfit minor differences. Therefore, the posterior distributions become very confident in the retrieved values. Even though the error bars from the pipeline have increased, we retrieve for an error inflation factor 10$^b$ \citep{Line2015} to account for model uncertainties in the 1D radiative transfer model and a potentially insufficient increase in the errors by the pipeline. 

We retrieve a smaller surface gravity and a slightly increased radius compared to the estimate of \citet{Zhang2025} and \citet{Kiman2025} based on self-consistent grid-models, while applying evolutionary model bounds to the priors. This then results in a significantly smaller mass than the previous estimates. This should highlight that obtained mass uncertainties in retrievals are not trustworthy. Again, the small posterior widths for the radius and the gravity lead to this effect.

The obtained abundances might as well be overconfident and thus, the isotope ratios we retrieve are likely overly constrained. For the oxygen ratios, we would expect slightly larger error estimates due to the lower SNR in the part of the spectrum, where the molecules are detected. However, their error bars are consistent with previous measurements \citep[e.g.][]{Gandhi2023}. 

\subsection{Disequilibrium chemistry} \label{discussion_dischem}
Based on Fig. \ref{fig:chem}, we observe CO clearly out of chemical equilibrium in COCONUTS-2b with abundances orders of magnitude larger than predicted by equilibrium chemistry. Thus, additional processes must be acting to modulate the chemical regime such that CO is present in the relatively cold atmospheric layers, for example, vertical mixing \citep{Zahnle2014}. 
The resulting vertical mixing strength we calculate (K$_{zz}$ of $\sim$10$^{3.3}$ cm$^2$/s) is compatible with quenching the equilibrium chemistry abundance of CO at a pressure of $\sim$40-50 bar. Using the Sonora Elf Owl models and the Gemini/FLAMINGOS-2 data \citet{Zhang2025} receive a stronger vertical mixing of  $\rm log(K_{zz})= 8.98^{+0.01}_{-0.03}$. \citet{Mukherjee2022} discuss the quench K$_{zz}$ dependent on the effective temperature of various targets and find for $\rm T_{eff}\approx 500K$ a typical K$_{zz}$ of $\sim$10$^6$ cm$^2$/s. This is larger compared to what we find, however, their sample has larger surface gravities than COCONUTS-2b. Smaller gravities lead to a decrease in K$_{zz}$ as presented in \citet{Miles2020}. \citet{Mukherjee2022} show that reaching the here observed values for CO requires small surface gravities or large vertical mixing. Our retrieved low surface gravity of about 3.9 dex and a rather small vertical mixing of K$_{zz}=$3.3$\rm cm^2/s$ for both retrievals is compatible with this low-gravity scenario, even though the probed parameter space by \citet{Mukherjee2022} does not extend to such low gravities. 

In contrast to CO, we find a smaller NH$_3$ abundance compared to the chemical equilibrium in both the free and constrained retrievals. \citet{Zahnle2014} propose NH$_3$ as a potential tracer for surface gravity, as its abundances is largely insensitive to vertical mixing, in contrast to species such as CO and CO$_2$. The smaller NH$_3$ abundance is in line again with the relatively low surface gravity that we have retrieved. 

Some studies have shown degeneracies between metallicity and surface gravity as the latter impacts the PT structure \citep{Molliere2015,Zhang2021b,deRegt2025}. Abundance ratios, such as the nitrogen isotope ratios presented here, seem to be more robust against this degeneracy. Even though the overall metallicities for the free and the constrained retrieval differ, the $^{14}$N/$^{15}$N and C/O ratios for example stay compatible as shown in Fig. \ref{fig:bulk} and \ref{fig:iso_ratio}.
Future analysis on the NIRSpec/G395H data set (Copeland et al. in prep.) will provide additional insights on the abundances of CO and CO$_2$, since their strongest absorption lines are visible at around 4.2 and 4.4 µm and will thus enable tighter constraints on the metallicity and C/O ratio.

Further, our retrievals on the binned spectrum place constraints on the abundances of PH$_3$, despite the absence of clearly identifiable absorption features. Using the full resolution retrievals in Fig. \ref{fig:isotop_h2s} in the appendix, we compare the best-fit spectra when including or removing one molecule from the line list. For PH$_3$, a weaker effect on the continuum improves the fit, while the strong feature expected at 10.1 µm is not visible. Although the logarithmic Bayes factor for including PH$_3$ (7.03) suggests statistical support and the retrieved abundance is broadly consistent with equilibrium chemistry, we consider the detection of of PH$_3$ to be tentative due to the lack of a distinct absorption feature. PH$_3$ is a molecule that has received significant attention with JWST. It has a significant abundance in Jupiter \citep{Larson1977} (due to its about 3 times higher metallicity, \citet{Mahaffy2000}), and was expected to be easily visible in brown dwarf spectra \citep{Miles2020}. However, early JWST observations did not detect this feature \citep[e.g.][]{Beiler2023, Matthews2025, Vasist2025}, and in particular \citep{Beiler2024b} highlighted that chemical models tend to over-predict PH$_3$ abundance relative to observations of T- and Y-dwarfs with temperatures below 500K. To date, PH$_3$ has only been observed in WISE0855, in which it was constrained to have an abundance of one part per billion \citep{Rowland2024} and recently at a larger abundance in Wolf1130C \citep{Burgasser2025}. The retrieved statistics favoring PH$_3$ in COCONUTS-2b, alongside the absence of a clear feature, motivate further work on this mysterious molecule.

\subsection{Ammonia isotope ratio consistent with ISM value}
$^{15}$NH$_3$ is only observable in the atmospheres of cold objects, and we are able to measure it using medium to high resolution data. As shown in  Fig. \ref{fig:fullvslowres}, the small features of the isotopologue are only discernible when using the full spectral resolution. This is in line with what \citet{Matthews2025} found by comparing the features of NH$_3$ and $^{15}$NH$_3$ for different temperature regimes: the warmer the targets the less evident the features, with a transition occurring at around 500K. Thus, COCONUTS-2b is just cold enough for a successful $^{15}$NH$_3$ detection. So far, COCONUTS-2b is the warmest object, where we could detect $^{15}$NH$_3$ and thus we demonstrate that we are able to uncover even such faint features using the full resolution retrievals. For the colder objects, WISE0855 and WISE1828, $^{15}$NH$_3$ was detected already in binned spectra of spectral resolution $\frac{\lambda}{\Delta \lambda} = 1000$.
We find the isotope ratio of $^{14}$N/$^{15}$N for COCONUTS-2b to be compatible with the ISM value within one to two sigma for the constrained and free retrievals. Given that COCONUTS-2b is a relatively young system, an isotope ratio close to the ISM value is consistent with formation from a molecular cloud with an average composition of the ISM. However, the retrieved ratio is also compatible within one to two sigma with the only two existing measurements of Y dwarfs \citep{Barrado2023, Kuhnle2025}, which are assumed to be either older due to their low effective temperatures or have lower masses as they would cool faster. This shows the need to measure more isotope ratios in a larger sample of brown dwarfs in the future to identify possible trends with age, temperature or other bulk parameters. 

\subsection{Heavy oxygen enrichment?}
We detect the oxygen isotopes $^{18}$O and $^{17}$O in water for the first time in a substellar companion's atmosphere. The retrieved ratios between the most common isotope $^{16}$O and the rarer (and heavier) isotopes indicate a significant enrichment in the rarer isotopes. Various processes have been discussed in the literature potentially leading to such an enrichment.

In case that COCONUTS-2b formed in a disk, a possible process would be photochemical dissociation in the disk. The more abundant isotopologue gets dissociated less likely due to self-shielding in the presence of incoming radiation of the host star. This leads to an enrichment of the less abundant, heavier isotopologues in the gas phase, which could subsequently be inherited by a gas giant forming in the disk. This process has been proposed for the planetary-mass object VHS-1256b \citep{Gandhi2023}. Such a mechanism requires a disk-formation pathway followed by outward migration. However, \citet{Ciesla2025} suggests that this process (isotope-selective photodissociation) in a disk is unlikely to produce oxygen isotope fractionation at levels that would be detected in the atmospheric composition of a companion forming from the disk.

Processes inside the atmosphere are unlikely explanations for the enrichment. On Earth, fractionation of heavier water isotopologues occurs in the hydrological cycle due to differentiation through condensation and evaporation \citep{Gat1996}. While condensation of water may occur in the cool atmosphere of COCONUTS-2b, the values of enrichment would be several orders of magnitude smaller than the values we received. 

Alternatively, COCONUTS-2b might have formed from a molecular cloud that was itself enriched in heavy oxygen isotopes. In this scenario, both the companion and its host star would be expected to share similar isotope ratios. Future high-precision observations of the host star's oxygen isotope composition will therefore be crucial for testing this hypothesis and constraining the formation pathway of the companion. 

Finally, it is possible that our retrieved values underestimate the oxygen isotope ratios in the atmosphere. Unaccounted atmospheric properties, such as clouds or the presence of additional molecules, could introduce degeneracies in the retrieved ratios. Atmospheric inhomogeneity, such as patchy clouds or hot spots, could further update the retrieved parameters. In fact, models that account for inhomogeneities predict different parameters compared to homogeneous models in a study on another planetary-mass brown dwarf \citep{ZJ2025b}. Further investigations on the model improvements are beyond the scope of this paper, but will be important to address in future studies. Nevertheless, the enrichment appears robust across different PT parameterizations and the values for $\rm ^{18}O/^{17}O$ are consistent with the ISM and previous observations. We are only now able to measure oxygen ratios in cold gas giant and brown dwarf atmospheres offering a new avenue for models to interpret enrichment. It is crucial to provide more observational measurements in the future to benchmark them.

\subsection{Sensitivity analysis on D/H} \label{dh}
Deuterium is especially interesting when discussing the mass estimate of a target. Brown dwarfs and gas giants are often distinguished by the deuterium burning limit at around 13 $\MJ$ \citep[e.g.][]{Burrows1997,Spiegel2011}. Below this mass gas giants are not able to burn deuterium to He and thus deuterium bearing isotoplogues would be present. Above this limit, objects are massive enough to burn deuterium and, depending on their mass, burn a large percentage of deuterium in the first hundreds of Myrs \citep{Spiegel2011}. Thus, we have tested for the presence of deuterium bearing species HDO and CH$_3$D but were unable to constrain the abundances of either of them. We nonetheless find for both upper limits beyond which the retrieval rules out the presence. For HDO the posterior distribution peaks before it falls off at the upper limit, hinting tentatively towards the presence of this molecule in the COCONUTS-2b atmosphere. CH$_3$D in contrast, falls off directly as seen in Fig. \ref{fig:DH_enrich} in the appendix. 

CH$_3$D has been proposed to be present and detectable in brown dwarf atmospheres \citep{Morley2019}. This isotopologue was found with a proto-solar D/H ratio in the atmosphere of the coldest brown dwarf WISE0855 \citep{Rowland2024}, which was  interpreted as evidence for a mass below the deuterium burning limit. The reported value of \cite{Spiegel2011} for the ratio between deuterium and hydrogen D/H for the ISM is (2.3$\pm$0.2)$\times$10$^{-5}$. Although our data do not allow us to directly constrain D/H in COCONUTS-2b, we calculate the upper bound of the D/H ratio with the derived upper limits on HDO and CH$_3$D according to: \begin{equation*}
    \rm (D/H)_{CH_4} = \frac{\rm VMR(CH_3D)}{4\times VMR(CH_4)},  \\
    \rm (D/H)_{H_2O} = \frac{\rm VMR(HDO)}{2\times VMR(H_2O)}.
\end{equation*} 
The calculations enables us to test whether this upper limit could be consistent with the ISM value. With the equations above, our calculated 3$\sigma$ (99.7\% percentile) upper limits for the $\rm(D/H)_{CH_4}$ ratio are 1.42$\times$10$^{-5}$ for the constrained retrieval and 6.29$\times$10$^{-6}$ for the free case. For $\rm(D/H)_{H_2O}$ we obtain an upper limit of 6.76$\times$10$^{-5}$ for the constrained retrieval and 1.34$\times$10$^{-4}$ for the free case. In Fig. \ref{fig:DH_enrich} in the appendix, we show the ratio in comparison with the ISM value. Interestingly, the free retrieval constrains the $\rm(D/H)_{H_2O}$ value around the ISM value and thus the upper limit we calculated would be consistent with our expectation of being consistent with the ISM value. However, the constrained case for $\rm(D/H)_{H_2O}$ for gives values smaller than the ISM value and falls off just right before the ISM value. As the latter PT profile is more constrained, it might miss to fit small features that the free profile can, thus not reaching an upper upper limit for HDO. CH$_3$D seems to be even more difficult to constrain as both PT parameterizations result in unconstrained values for $\rm(D/H)_{CH_4}$ clearly excluding values smaller than $\sim$10$^{-5}$. The lower two panels in Fig. \ref{fig:DH_enrich} in the appendix show the spectral features from the retrieved case compared to a model excluding HDO or CH$_3$D. The spectral differences are subtle, suggesting that the expected features of those molecules lie close to the detection threshold with the MIRI/MRS data. While the current dataset does not have the SNR or spectral resolution to conclusively detect deuterium bearing isotopes, these results provide a tantalizing hint that deuterium may be present in the atmosphere. If present, this would confirm previous studies that have determined the mass to be in the planetary regime, below the deuterium-burning limit. Therefore, future observations of COCONUTS-2b reaching a higher SNR or spectral resolution are needed to conclude on the presence of the deuterium-bearing molecules.

\subsection{Non-detection of hydrocarbons} \label{discussion_w0458}
\begin{figure*}[h]
    \centering
    \includegraphics[width=\linewidth]{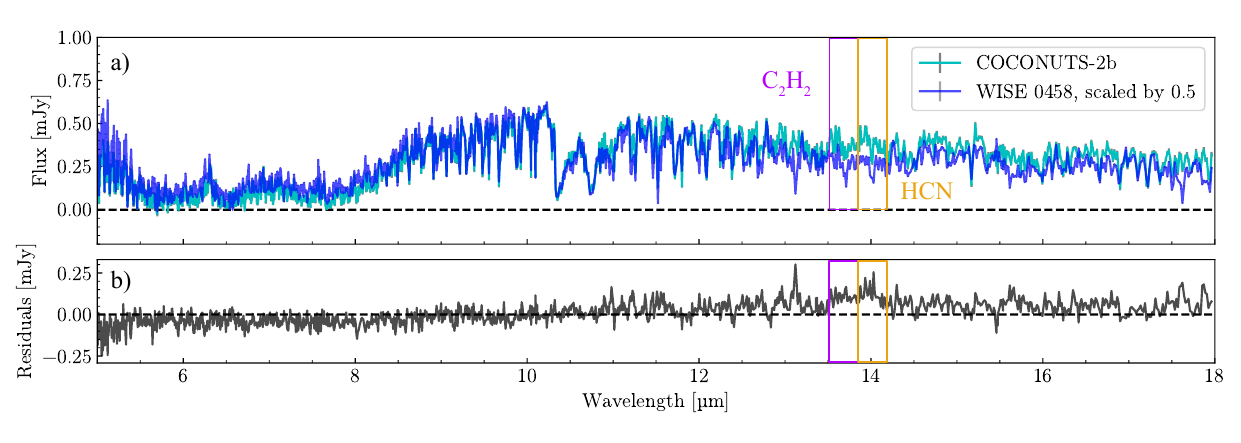}
    \caption{Comparison of the spectrum of COCONUTS-2b to WISE 0458 \citep{Matthews2025}. In panel a) both spectra are plotted on top of each other with the spectrum of WISE 0458 being scaled by a factor of 0.5. in panel b) the difference between COCONUTS-2b and WISE 0458 is shown. The purple and orange boxes indicate where in the spectrum the differences in absorption of C$_2$H$_2$ and HCN are located. We do not see the absorption of these molecules in COCONUTS-2b in spite of the presence in WISE 0458 \citep{Matthews2025} and their similar spectral type.}
    \label{fig:comparison}
\end{figure*}

WISE J045853.90+643451.9 (hereafter WISE0458) is a T8.5-T9.0 brown dwarf binary and was observed with MIRI/MRS with an unexpected detection of HCN and C$_2$H$_2$ \citep{Matthews2025}. Due to the similar spectral type and effective temperature of $\sim$500K, this brown dwarf binary is comparable to COCONUTS-2b; in fact up to now the only other T-dwarf observed with MIRI/MRS making COCONUTS-2b the first opportunity to study the production of these molecules in more detail. In Fig. \ref{fig:comparison} we overlay both acquired spectra, where we scale the flux of WISE0458 with a factor of 0.5. The lower panel shows the difference between the two spectra. As WISE0458 is slightly warmer in effective temperature compared to COCONUTS-2b, we see a slope in the residuals with negative residuals for shorter and positive for larger wavelengths, demonstrating that the compositions are generally similar. In WISE0458, clear absorption features of C$_2$H$_2$ and HCN are visible in the purple and orange boxes, respectively. We do not detect these molecules in COCONUTS-2b, which is indicated by the offset in the difference plot and confirmed by a corresponding retrieval analysis, where we do not detect the two molecules with a logarithmic Bayes factor of 0.19 and -0.72 for C$_2$H$_2$ and HCN, respectively.  

The presence of HCN and C$_2$H$_2$ has been discussed as a general property of T-dwarf atmospheres \citep{Matthews2025}.  Models predict HCN in cold atmospheres to be connected to strong vertical mixing and high surface gravities \citep{Zahnle2014}. As we find a rather small value for the vertical mixing parameter (K$_{zz}=3.3$) and gravity (logg$\approx$3.9 dex), a non-detection is plausible.
Explaining the presence of C$_2$H$_2$ in WISE0458 indicates an incomplete chemical network and/or more complex and unaccounted atmospheric processes \citep{Matthews2025}. The non-detection in COCONUTS-2b being in line with current models adds up to the complexity and shows the diversity in compositions for T-type objects. Thus, we can rule out their presence being strictly spectral type dependent. Instead other characteristics, such as the surface gravity, need to be taken into account. This is another indication that our current understanding of the processes leading to the disequilibrium chemistry in such cold atmospheres is incomplete. More data in the near and MIR for late T- and Y-dwarfs will help to improve our understanding of carbon and nitrogen chemistry in these cold atmospheres.

\subsection{A note on clouds} \label{discussion_clouds}
Salt clouds such as Na$_2$S and KCl have been proposed in objects at the T-/Y- transition \citep{Morley2012,Manjavacas2022}. However, in contrast to the strong silicate and iron cloud features in hotter L/T- transition objects \citep[e.g.][]{Suarez2022,Miles2023,Molliere2025}, there are no distinct spectral features of salt clouds in the observed wavelength range with MIRI/MRS.
Low lying silicate clouds have been proposed to change the available chemical budget through rain-out processes \citep{Calamari2024}. This has been shown to change the retrieved elemental abundances, such as C/O. To account for this, we apply the proposed correction factor to our stated C/O values. In contrast to pure absorption features, scattering from cloud particles may still affect the retrieved spectrum. However, in our analysis, we can already explain the overall flux distribution well without including any scattering effects from clouds. Especially, we are able to fit well both the Gemini/FLAMINGOS-2 and the MIRI/MRS parts together with a clear atmosphere model. We have explored various setups using cloudy retrievals including Na$_2$S, KCl and water clouds, however, they were not conclusive. The strongest variations in the resulting fits were located between 3 and 5 µm, where we do not include any data in this analysis. Future studies, ideally ranging over the full spectral energy distribution, may focus on whether condensing, potentially patchy, clouds still might be present. 

\subsection{Compositional comparison to the host star}
Few observations have been taken so far of the host star L34-26. Its metallicity is estimated to be solar, with a value of $\rm [M/H]=0.00\pm0.08$ \citep{Hojjatpanah2019}, while no measurements of its C/O ratio have been reported. Future observations would be needed to constrain the elemental and isotope abundances of the M dwarf, enabling comparative studies between the host star and the companion. In principle, similar elemental and isotopic ratios would indicate a potentially binary-system-like formation scenario, in which both the host and companion formed from the same molecular cloud. Conversely, significant differences in elemental or isotopic compositions could point to alternative formation pathway scenarios, such as disk-related processes or gravitational capture \citep{Marocco2024}.
Future observations of cold T- and Y- brown dwarfs using MIRI/MRS and NIRspec, as planned in JWST programs GO 3647 (PI: P. Patapis), GO 5765 (PI: E. Matthews) and GO 8441 (PI: J. Faherty) will provide crucial comparative reference measurements for elemental and isotopic ratios in substellar atmospheres. Extending similar analyses to additional cold systems, such as Eps Ind, TWA 7 and Her 14c, will further improve our understanding of the diversity of formation pathways of cold companions.
Finally, ground-based telescopes, such as the future ELT, will provide even higher resolution measurements on these faint companions along with their host stars on smaller wavelength coverages of the spectrum. 

\section{Summary and outlook} \label{conclusion} 
We present the MIRI/MRS observations of the cold ($\rm T_{eff}\approx 480 K$) far-out planetary-mass companion COCONUTS-2b (Program ID: 6463, PI: P. Patapis) with an exceptional SNR of up to 40 at 11.8µm. To study its chemical composition, we perform a comprehensive atmospheric retrieval analysis of the full resolution of the spectrum. So far, retrievals of MIRI/MRS data have been performed on binned spectra using a spectral resolution of R$\sim$1000. The higher resolution allows us to detect nitrogen and oxygen isotopologues and allows for a more detailed analysis of the composition of this atmosphere than possible in previous studies. In our setup, we choose two different PT parametrization, either a free or a constrained one and summarize both results in the following:
\begin{itemize}
    \item We detect the molecules H$_2^{18}$O and H$_2^{17}$O and constrain the ratios of $^{16}$O/$^{18}$O to be 247$^{+27}_{-24}$ and 256$^{+29}_{-25}$, $^{16}$O/$^{17}$O to be 662$^{+98}_{-83}$ and 934$^{+174}_{-139}$ and $^{18}$O/$^{17}$O 2.7$^{+0.6}_{-0.4}$ and 3.7$^{+0.8}_{-0.7}$ for the free and constrained retrieval respectively, indicating an enrichment in the heavier oxygen isotopes. 
    \item We detect the nitrogen isotopologue $^{15}$NH$_3$ and constrain the ratio to be 337$^{+47}_{-38}$ and 324$^{+46}_{-40}$ for the free and constrained profile respectively. They are within one to two sigma compatible with the value we found for the ISM and previous Y dwarf observations.
    \item We do not detect HCN and C$_2$H$_2$ and only detect PH$_3$, CO$_2$, and $^{13}$CH$_4$ tentatively using the Bayes factor comparison. 
    \item As we do not constrain HDO and CH$_3$D, but obtain only upper limits for both species, we likely are at the sensitivity limit to detect deuterium-bearing molecules. 
    \item We find a subsolar to solar metallicity and a subsolar C/O ratio compatible with previous measurements.
\end{itemize}

The measured isotopologue abundances will be important when comparing to the host stars composition in order to study the formation of COCONUTS-2b. In addition, future data with even higher SNR or higher spectral resolution will potentially result in more accurate ratios and might allow us to detect deuterium bearing species - if present. This may provide an additional constraint on the mass estimate for this wide-separation companion, whose dynamical mass will be difficult to obtain. 
An atmospheric retrieval analysis using the NIRSpec data set will potentially lead to a better understanding of the vertical mixing due to better constraints on CO and CO$_2$. Further comparisons using self-consistent grid models, such as the study done on MIRI/LRS data, will be crucial for a broader characterization of COCONUTS-2b (Ravet et al., in review). Combining the NIRSpec, MIRI/LRS and MIRI/MRS data sets will ultimately allow us to  include more complex features in our models, such as adding clouds or vertically changing abundance profiles.
Thus, the atmosphere of cold COCONUTS-2b still holds many mysteries waiting to be solved. 

\section*{Data availability}
The raw data of the MIRI/MRS data set used in this analysis is available on MAST (\url{https://mast.stsci.edu/}) under the PID: 6463 or directly through the DOI: \url{https://doi.org/10.17909/12mr-hg42}. The reduced data will be made available under the following link: \url{https://doi.org/10.5281/zenodo.19887581}.

\begin{acknowledgements}
We thank the anonymous referee for the comments that improved the quality of this paper. We want to thank Céline Nussbaumer and Jérémie Pierre for their insights on the retrieval setups through their semester projects and Jean Hayoz and Henrik Knierim for fruitful discussions. Further, we would like to thank Fabian Grübel for valuable additions.
PP thanks the Swiss National Science Foundation (SNSF) for financial support under grant number 200020\_200399.
DB is supported by Spanish MCIN/AEI/10.13039/501100011033 grant PID2023-150468NB-I00 and No. MDM-2017-0737.
NW acknowledges funding from NSF award \#2238468, \#1909776, and NASA Award \#80NSSC22K0142. NW acknowledges support for program \#06463 was provided by NASA through a grant from the Space Telescope Science Institute, which is operated by the Association of Universities for Research in Astronomy, Inc., under NASA contract NAS 5-03127. MB acknowledges support in France from the French National Research Agency (ANR) through project grants ANR-20-CE31-0012 and ANR-23-CE31-0006 and the Action Thématique Exosystèmes as part of the CNRS/INSU.
\end{acknowledgements}

\bibliographystyle{aa} 
\bibliography{export} 

\begin{appendix}
\section{Observation}
\begin{figure*}[h]
    \centering
    \includegraphics[width=\linewidth]{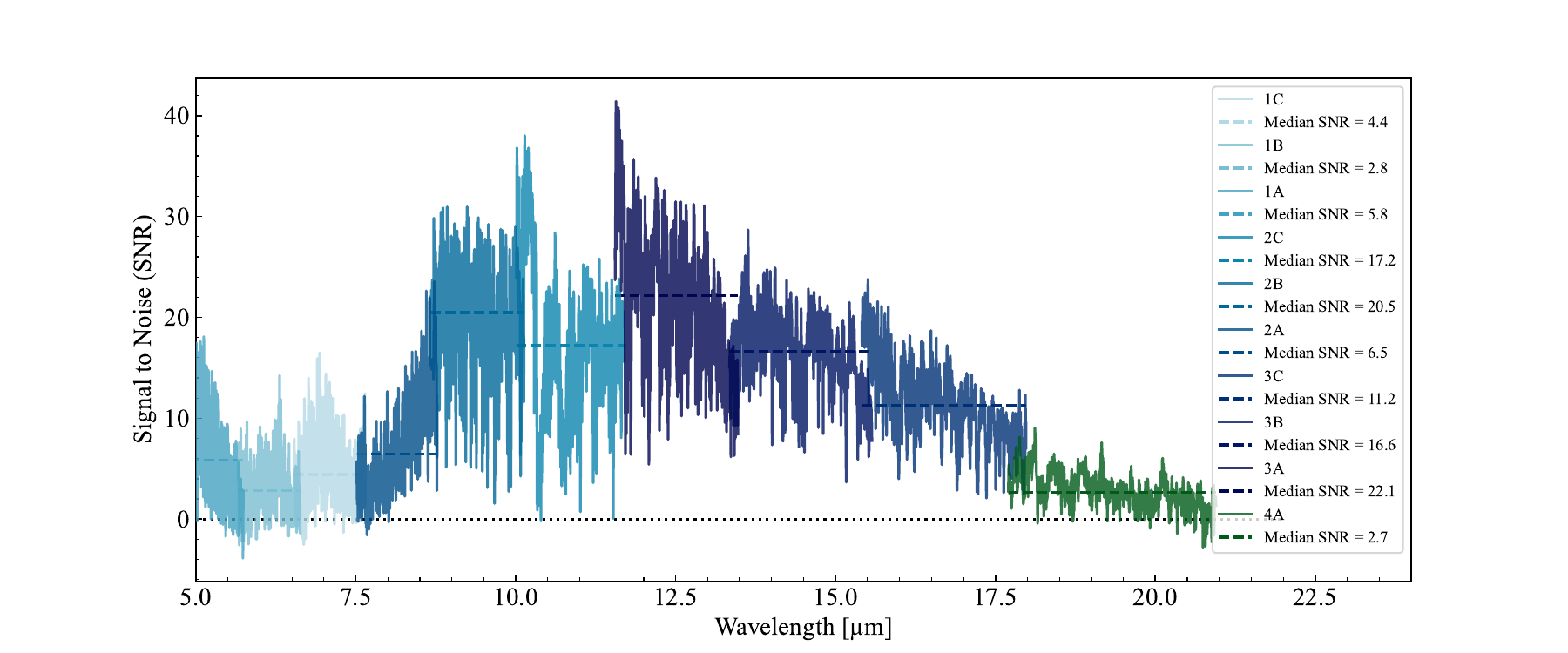}
    \caption{Signal to noise ratio per channel for the MIRI/MRS observation with median values highlighted in dashed lines and presented in the legend.}
    \label{fig:app_snr}
\end{figure*}

\begin{figure*}[h]
    \centering
    \includegraphics[width=\linewidth]{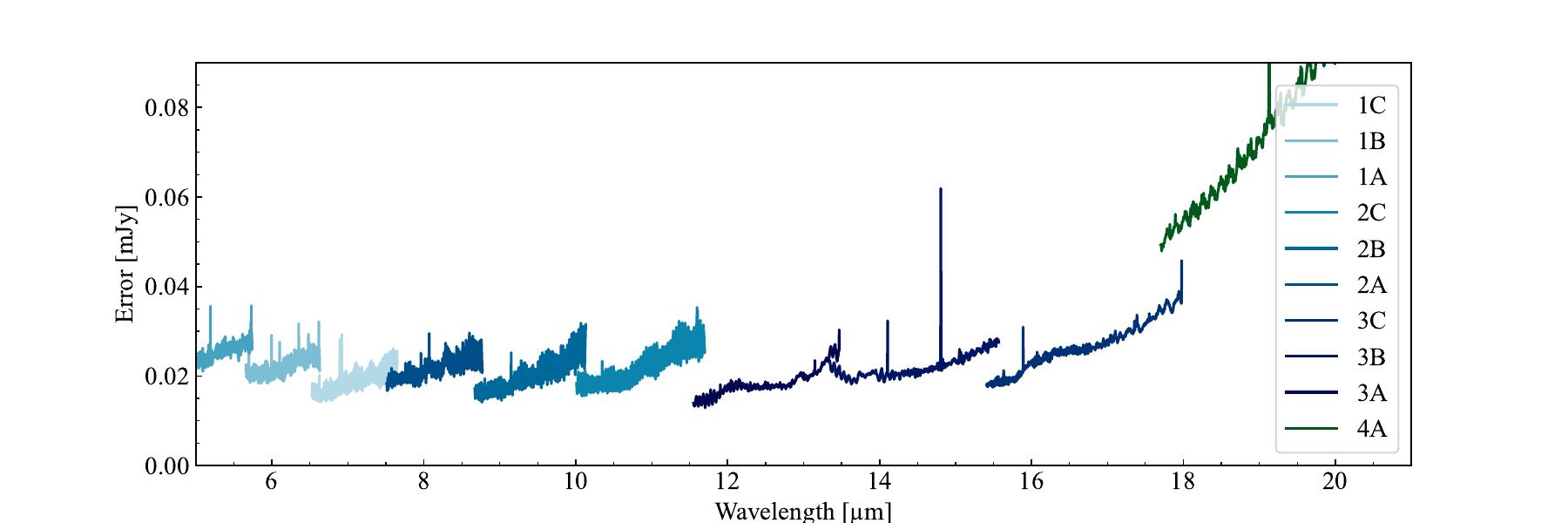}
    \caption{Error from the pipeline across the wavelengths.}
    \label{fig:app_err}
\end{figure*}

We present the Signal to Noise (SNR) for the resulting observation form 5 to 21 µm in Fig. \ref{fig:app_snr}. The maximum median SNR across the channels was reached in channel 3A with 22.1 reaching up to $\sim$40. The lowest was reached in channel 4A with a value of 2.7. Thus, we do not include the latter in the analysis.
In Fig. \ref{fig:app_err} we present the errors on the data set in mJy. The largest errors are found in channel 4A and 3C after that. The lowest can be found in channel 3A. 

\section{Retrievals}
\begin{figure*}[h]
    \centering
    \includegraphics[width=\linewidth]{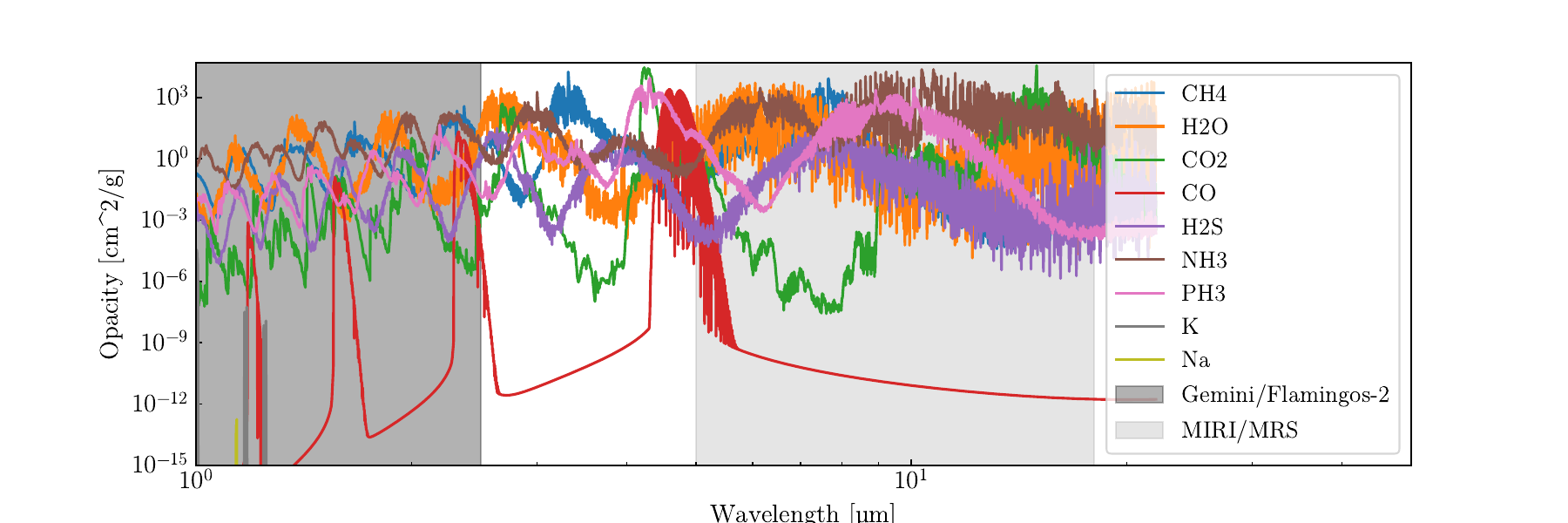}
    \caption{Opacities used for the retrievals at $\frac{\lambda}{\Delta \lambda} = 1000$.}
    \label{fig:app_opacities}
\end{figure*}

\begin{figure*}[h]
    \centering
    \includegraphics[width=\linewidth]{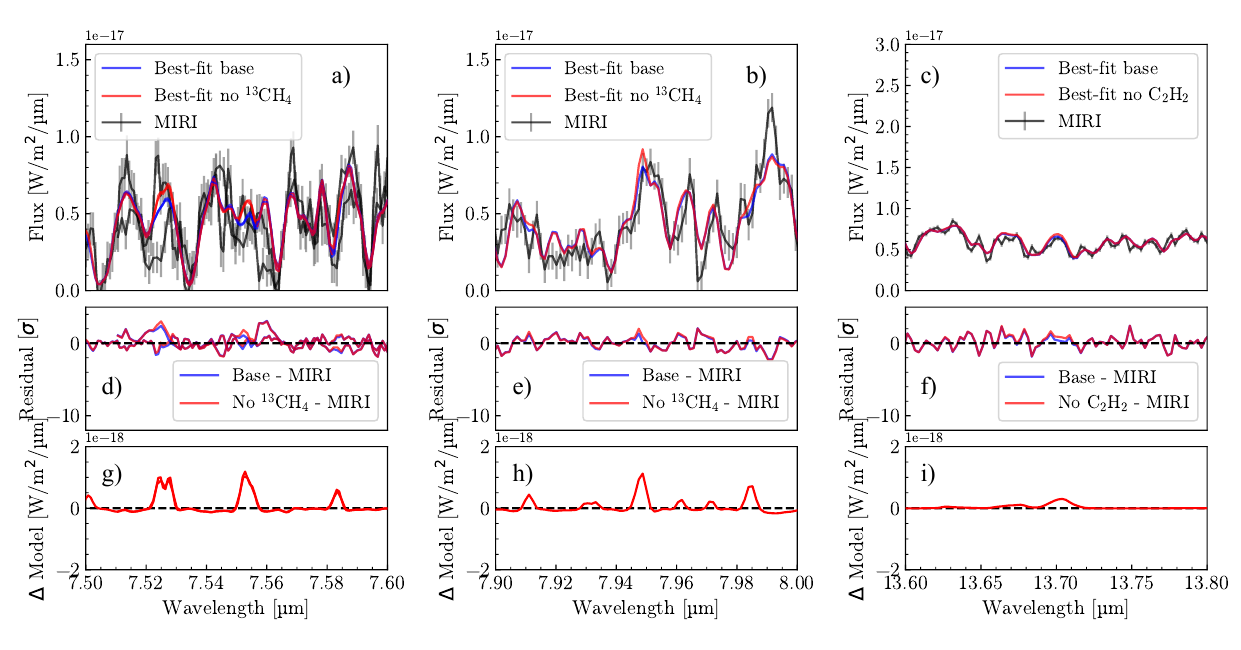}
    \caption{Best-fit retrievals on full MIRI/MRS resolution with and without including $^{13}$CH$_4$ and C$_2$H$_2$ in blue and red respectively compared to the data in black with the same panel structure as in Fig. \ref{fig:isotop_15nh3}. Here we show the part of the spectrum with the normalized largest difference between the best-fits.}
    \label{fig:isotop_13ch4}
\end{figure*}

\begin{figure*}[h]
    \centering
    \includegraphics[width=\linewidth]{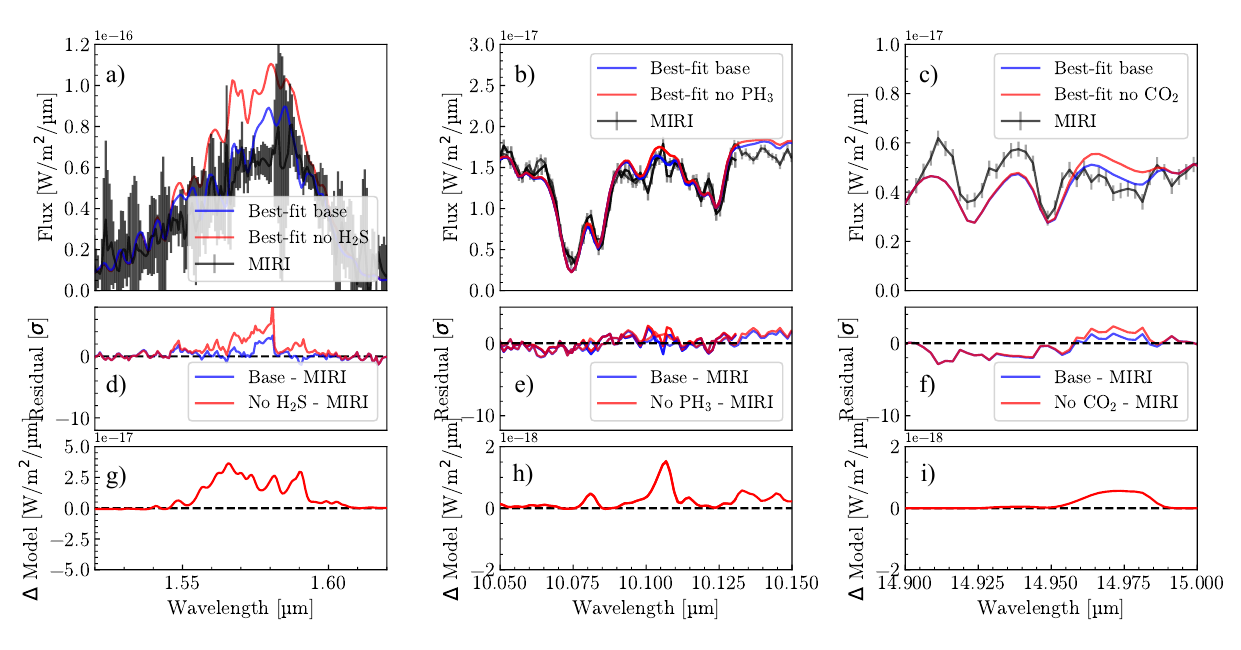}
    \caption{Best-fit retrievals on full MIRI/MRS resolution with and without including H$_2$S, PH$_3$ and CO$_2$ in blue and red respectively compared to the data in black with the same panel structure as in Fig. \ref{fig:isotop_15nh3}. Here we show the part of the spectrum with the normalized largest difference between the best-fits.}
    \label{fig:isotop_h2s}
\end{figure*}

\begin{figure*}[h]
    \centering
    \includegraphics[width=\linewidth]{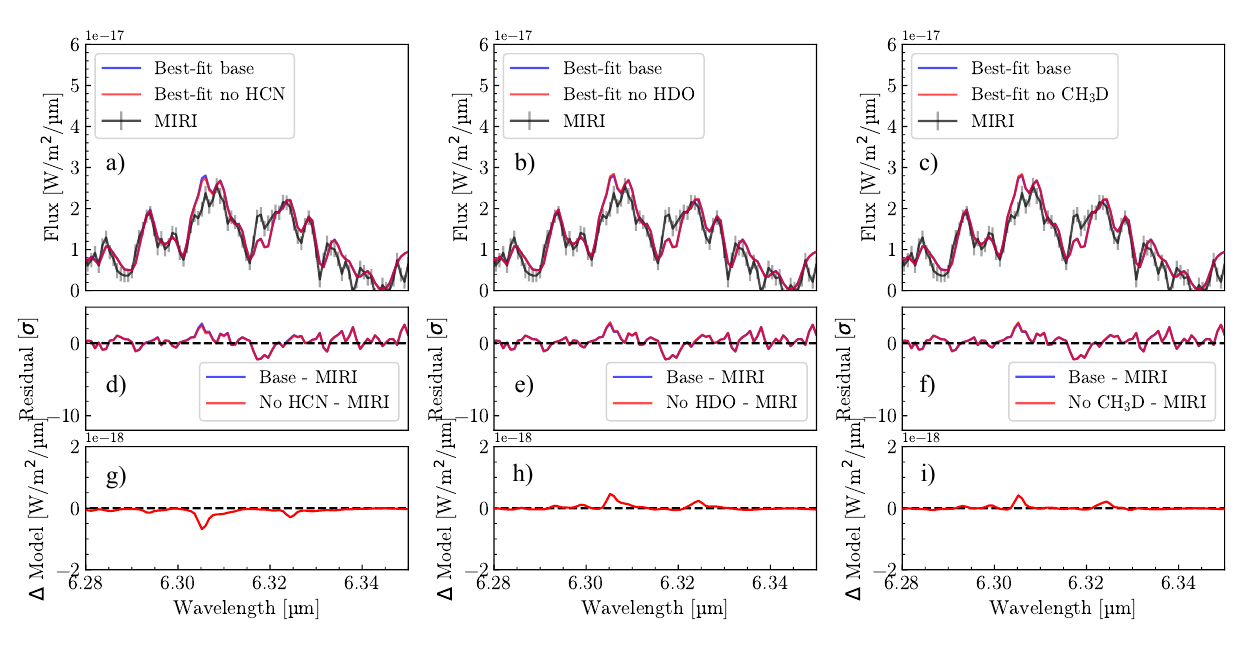}
    \caption{Best-fit retrievals on full MIRI/MRS resolution with and without including HCN, HDO and CH$_3$D in blue and red respectively compared to the data in black with the same panel structure as in Fig. \ref{fig:isotop_15nh3}. Here we show the part of the spectrum with the normalized largest difference between the best-fits.}
    \label{fig:isotop_hcn}
\end{figure*}

\begin{figure*}[h]
    \centering
    \includegraphics[width=\linewidth]{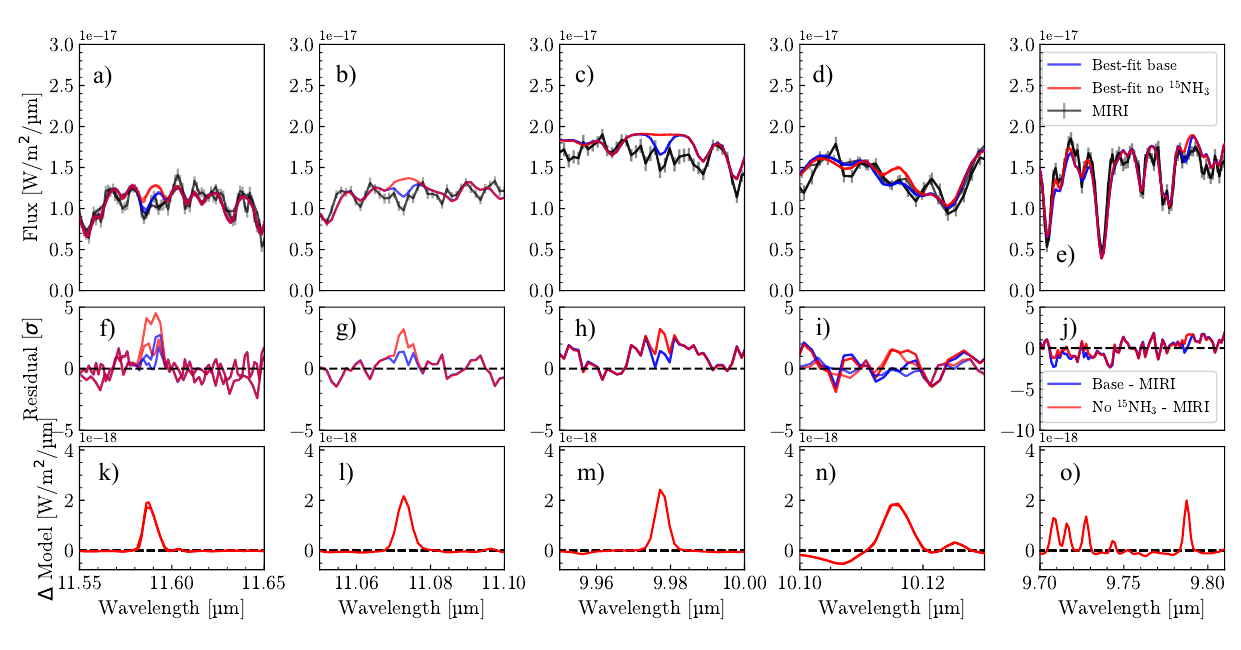}
    \caption{Best-fit retrievals on full MIRI/MRS resolution with and without including ${15}$NH$_3$ in blue and red respectively compared to the data in black with the same panel structure as in Fig. \ref{fig:isotop_15nh3}. Here we show the part of the spectrum with the fourth to eight largest differences between the best-fits.}
    \label{fig:isotop_15nh3_all}
\end{figure*}

\begin{figure*}[h]
    \centering
    \includegraphics[width=\linewidth]{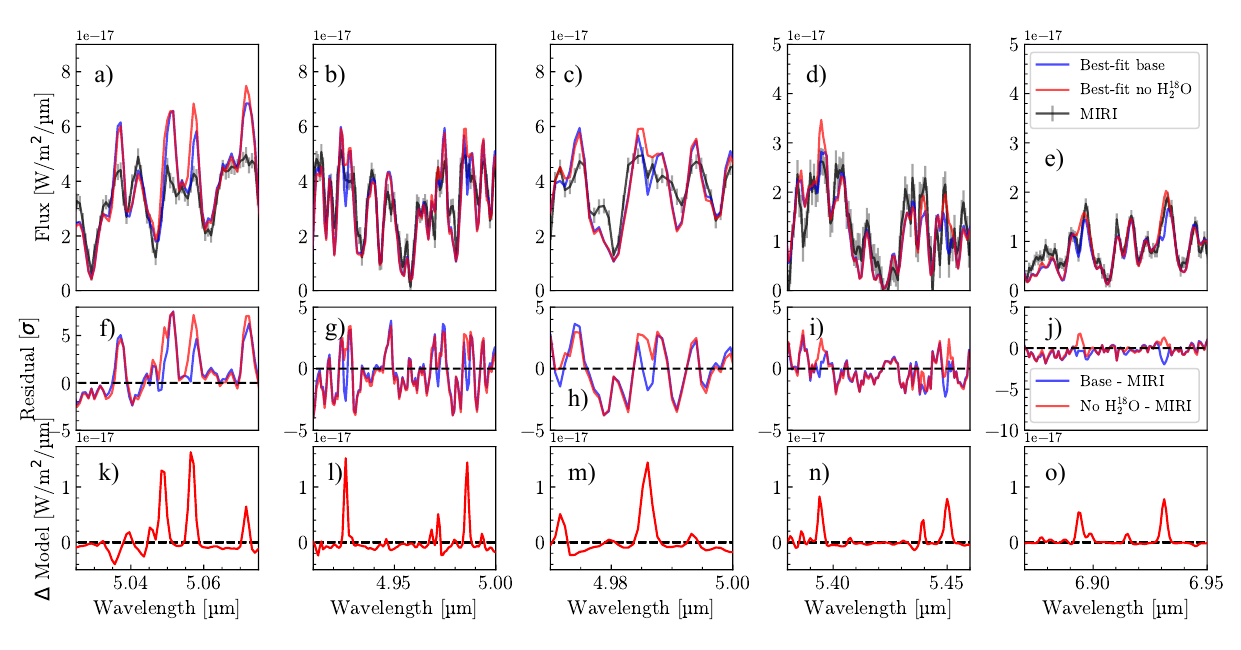}
    \caption{Best-fit retrievals on full MIRI/MRS resolution with and without including H$_2^{18}$O in blue and red respectively compared to the data in black with the same panel structure as in Fig. \ref{fig:isotop_h218o}. Here we show the part of the spectrum with the largest, second, third, seventh and eight largest differences between the best-fits.}
    \label{fig:isotop_h218o_all}
\end{figure*}

\begin{figure*}[h]
    \centering
    \includegraphics[width=\linewidth]{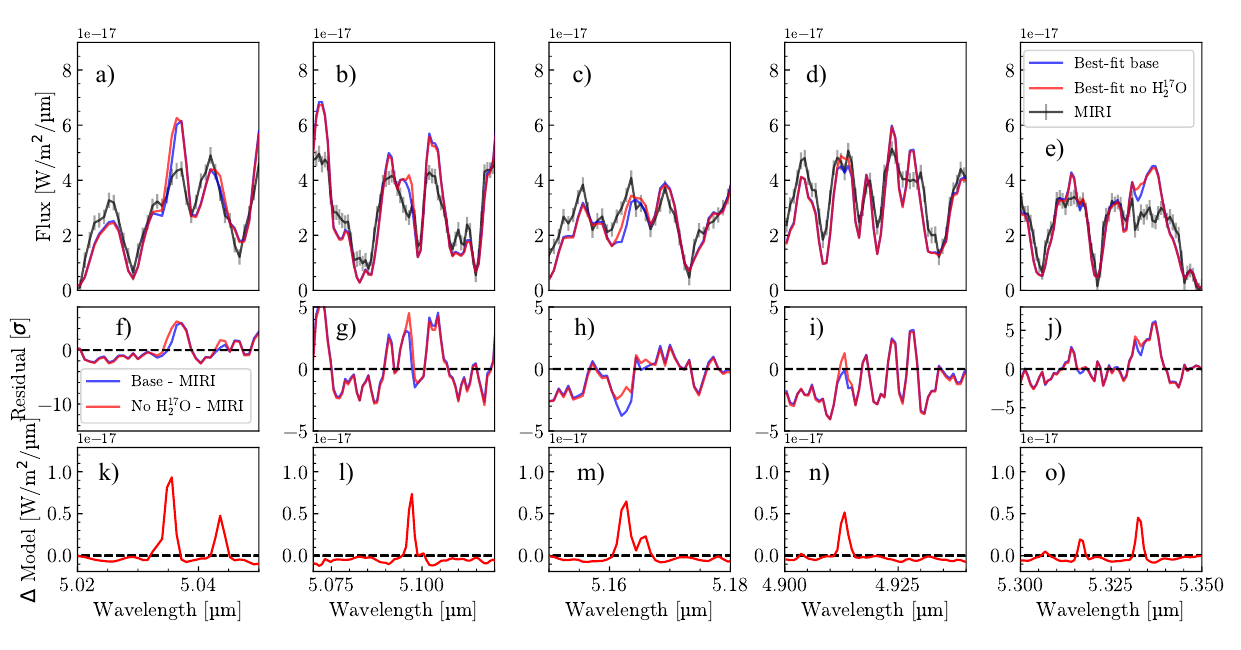}
    \caption{Best-fit retrievals on full MIRI/MRS resolution with and without including H$_2^{17}$O in blue and red respectively compared to the data in black with the same panel structure as in Fig. \ref{fig:isotop_h217o}. Here we show the part of the spectrum with the fourth to eight largest differences between the best-fits.}
    \label{fig:isotop_h217o_all}
\end{figure*}

\begin{figure*} 
    \centering
    \includegraphics[width=\linewidth]{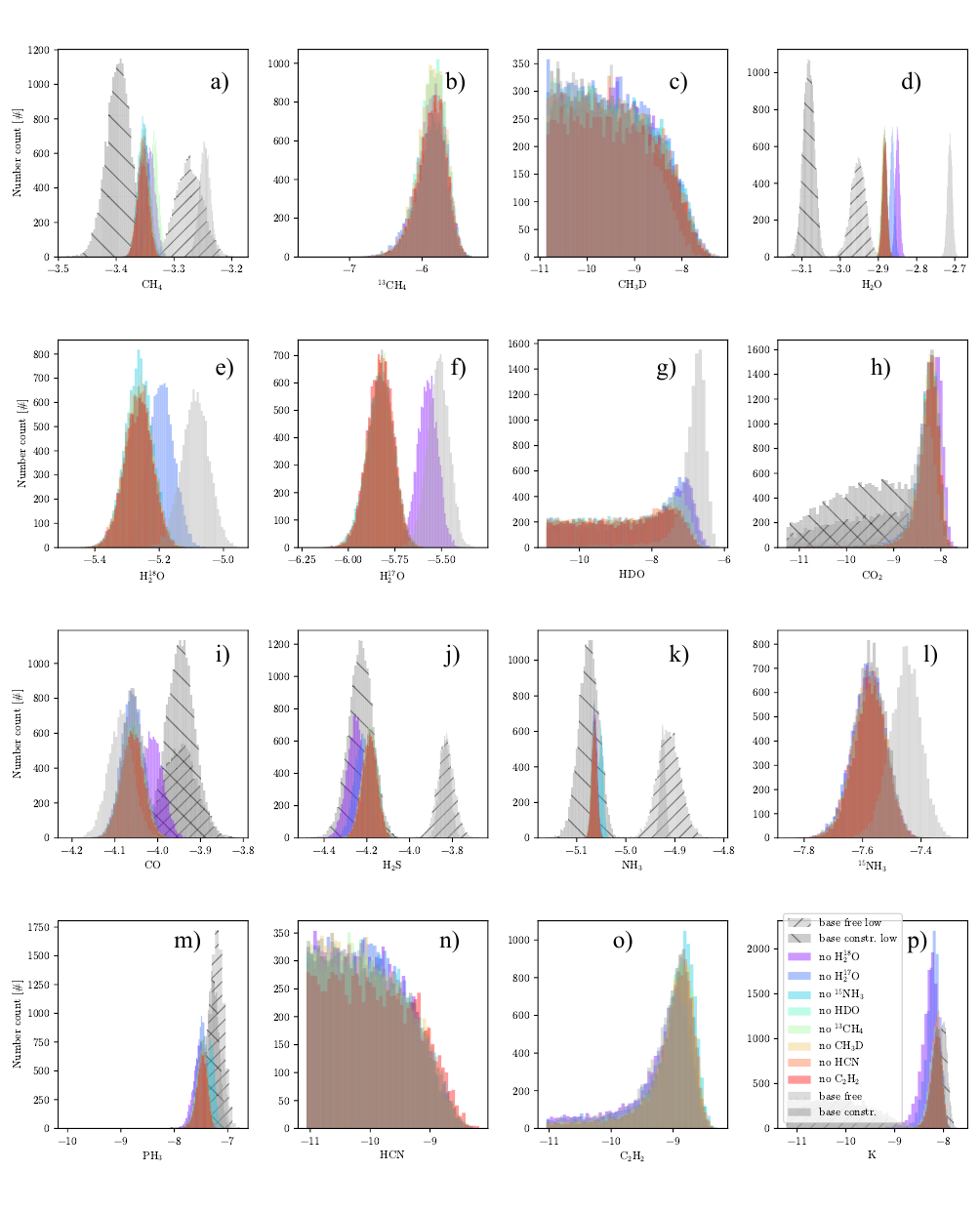}
    \caption{Abundances in comparison between the full and the low resolution as well as the base and the leave-one-out retrievals.}
    \label{fig:app_abund_fullvslowres}
\end{figure*}

\begin{figure*} [h]
    \centering
    \includegraphics[width=0.5\linewidth]{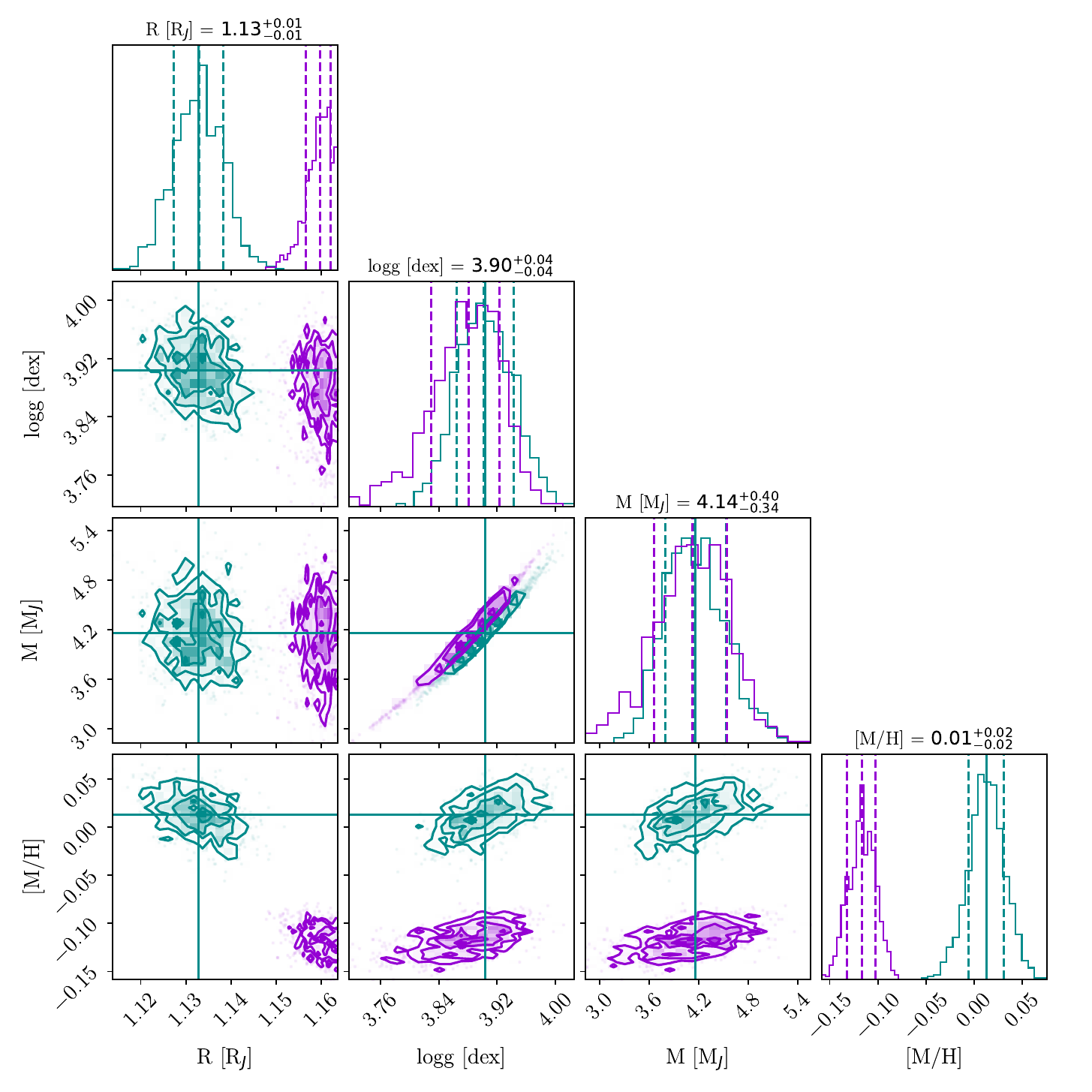}
    \caption{Corner plot of the retrieved and calculated bulk parameters for the free and the constrained case in green and violet, respectively: the radius, logg, mass and metallicity. These results are based on the low resolution retrievals}
    \label{fig:app_corner}
\end{figure*}

\begin{figure*}[h]
    \centering
    \includegraphics[width=0.75\linewidth]{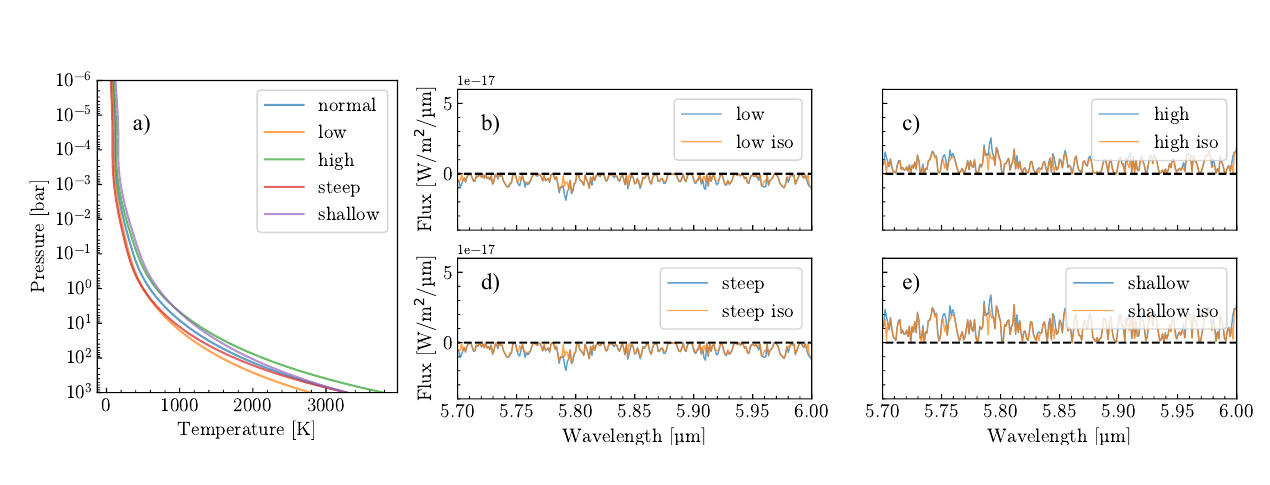}
    \caption{Effect on the model when varying the PT profile. Panel a) shows the variation of the PT profile for either a shift or a change in the slope of the PT profile. The effect of on the spectrum either including ('iso') or neglecting the isotopologue is shown in panels b)-e). In b) we shift the PT profile by 500K towards lower temperatures, c) a shift by 500K towards higher temperatures, d) a steeper profile by multiplying each node by 10\% and e) a shallower profile by 10\%.}
    \label{fig:app_iso_pt} 
\end{figure*}

\begin{figure*} 
    \centering
    \includegraphics[width=0.6\linewidth]{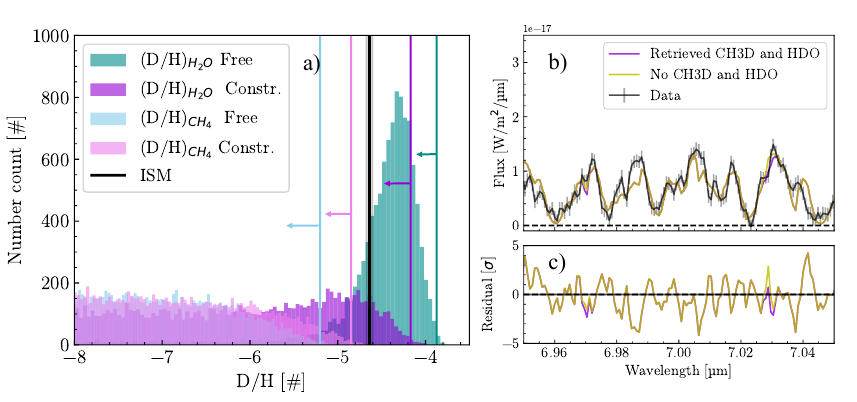}
    \caption{The calculated $\rm (D/H)_{H_2O}$ ratio from the upper limits for the free and constrained retrieval in green and violet and the ratio $\rm (D/H)_{CH_4}$ for the free and constrained case in blue and pink compared to the ISM value \citep{Spiegel2011} in black in panel a). Panel b) presents the modeled spectrum without HDO and CH$_3$D in yellow, the retrieved best-fit in violet including the two molecules compared to the data. The residuals to plot b) are shown in panel c).}
    \label{fig:DH_enrich}
   
\end{figure*}

In this appendix we provide several supporting plots for the retrieval analysis. We present in Fig. \ref{fig:app_opacities} the opacities used in c-k resolution. We also highlight the Gemini/FLAMINGOS-2 and MIRI/MRS covered wavelength ranges.
We present the median and 16th and 84th percentiles of the posteriors of the low resolution retrievals in Tab. \ref{tab:ret_val}.
The full resolution retrieval comparisons for species not clearly detected are shown for $^{13}$CH$_4$ and C$_2$H$_2$ in Fig. \ref{fig:isotop_13ch4}, for H$_2$, PH$_3$ and CO$_2$ in Fig. \ref{fig:isotop_h2s} and HCN, HDO and CH$_3$D in Fig. \ref{fig:isotop_hcn}. Here, we present the areas in the spectrum where the difference in the model is the strongest. In Fig. \ref{fig:isotop_15nh3_all}, \ref{fig:isotop_h218o_all},  and \ref{fig:isotop_h217o_all} we present more wavelength areas in the spectrum compared to the ones presented in Fig. \ref{fig:isotop_h218o},\ref{fig:isotop_h217o}, and \ref{fig:isotop_15nh3}, where the absorption of $^{15}$NH$_3$, H$_2^{18}$O and H$_2^{17}$O are visible, respectively. The comparison of all retrieved molecular abundances including the ones from the low and full resolution retrievals are summarized in Fig. \ref{fig:app_abund_fullvslowres}. In Fig \ref{fig:app_corner} we present the correlations between some of the bulk parameter retrieved and calculated from the low resolution retrieval results. 
Fig. \ref{fig:app_iso_pt} shows how the variations in the PT profile vary the spectrum either including the isotopologues or not. We present four cases, where we either increase or decrease the bottom temperature or vary the slopes to either a stepper or shallower profile. In Fig. \ref{fig:DH_enrich}, we present the $(\rm D/H)_{H_2O}$ and $(\rm D/H)_{CH_4}$ ratio for the free and constrained case and compare them to the ISM value.

\begin{table}[h]
    \begin{center}
    \caption{Retrieved and calculated (marked with *) parameter for the free and constrained cases of the low resolution retrievals. }
    \small
    \label{tab:ret_val}
    \begin{tabular}{l|l|l}
         Value &  Free retrieval & Constrained retrieval \\ \hline \hline
         Radius [$\RJ$] & 1.133$^{+0.005}_{-0.006}$ [1.156] & 1.160$^{+0.002}_{-0.003}$ [1.189] \\
         log g [dex] & 3.90$^{+0.04}_{-0.04}$ [3.89] & 3.88$^{+0.04}_{-0.05} [4.01]$\\
         mass [$\MJ$] (*) & 4.1$^{+0.4}_{-0.3}$ & 4.1$^{+0.4}_{-0.5}$\\
         $\rm T_{eff}$ [K] (*) & 471.3$^{+1.6}_{-1.4}$ & 468.1$^{+1.1}_{-1.1}$\\
         $\rm T_{deep}$ [K] &  3895$\pm$593 [3017] & 3428$\pm$91 [3290] \\
         T$_{1}$ [\#] & 0.43$\pm$0.05 [0.52] & - \\
         T$_{2}$ [\#] & 0.70$\pm$0.03 [0.74] & - \\
         T$_{3}$ [\#] & 0.552$\pm$0.002 [0.554] & - \\
         T$_{4}$ [\#] & 0.630$\pm$0.004 [0.626] & - \\
         T$_{5}$ [\#] & 0.32$\pm$0.04 [0.35] & - \\
         T$_{6}$ [\#] & 0.8$\pm$0.3 [0.9] & - \\
         T$_{7}$ [\#] & 0.7$\pm$0.3 [0.5] & - \\
         T$_{8}$ [\#] & 0.7$\pm$0.3 [0.3] & - \\
         T$_{9}$ [\#] & 0.7$\pm$0.3 [0.2] & -\\
        
         dlogT/dlogP$_{1}$ [\#] & - & 0.24$\pm$0.01 [0.25] \\
         dlogT/dlogP$_{2}$ [\#] & - & 0.251$\pm$0.0051 [0.246] \\
         dlogT/dlogP$_{3}$ [\#] & - & 0.245$\pm$0.003 [0.249] \\
         dlogT/dlogP$_{4}$ [\#] & - & 0.248$\pm$0.004 [0.238] \\
         dlogT/dlogP$_{5}$ [\#] & - & 0.181$\pm$0.004 [0.180] \\
         dlogT/dlogP$_{6}$ [\#] & - & 0.23$\pm$0.02 [0.22] \\
         dlogT/dlogP$_{7}$ [\#] & - & 0.16$\pm$0.04 [0.18] \\
         dlogT/dlogP$_{8}$ [\#] & - & 0.03$\pm$0.04 [0.00] \\
         dlogT/dlogP$_{9}$ [\#] & - & 0.02$\pm$0.04 [0.08] \\
         dlogT/dlogP$_{10}$ [\#] & - & 0.03$\pm$0.04 [0.10] \\
         
         $\rm [M/H]$ [\#] (*) &  0.01$^{+0.02}_{-0.02}$ & -0.12$^{+0.01}_{-0.02}$\\
         C/O [\#] (*) & 0.43$^{+0.02}_{-0.01}$& 0.45$^{+0.02}_{-0.02}$\\
         H$_2$ [\#] & 0.83840$^{+0.00009}_{-0.0001}$ & 0.83884$^{+0.00005}_{-0.00004}$ \\
         He [\#] & 0.15969$^{+0.00002}_{-0.00002}$ & 0.159779$^{+0.000009}_{-0.000008}$\\
         log VMR CH$_4$ [\#] & -3.28$^{+0.03}_{-0.03}$ & -3.42$^{+0.02}_{-0.02}$ \\
         log VMR H$_2$O [\#] & -2.96$^{+0.03}_{-0.03}$ & -3.09$^{+0.02}_{-0.02}$ \\
         log VMR CO$_2$ [\#] & -9.5$^{+0.9}_{-1.0}$& -9.6$^{+0.9}_{-0.9}$\\
         log VMR CO [\#] & -3.95$^{+0.05}_{-0.05}$ & -3.94$^{+0.04}_{-0.05}$\\
         log VMR H$_2$S [\#] & -3.84$^{+0.05}_{-0.05}$ & -4.25$^{+0.08}_{-0.08}$\\
         log VMR NH$_3$ [\#] & -4.91$^{+0.03}_{-0.03}$ & -5.09$^{+0.02}_{-0.02}$\\
         log VMR PH$_3$ [\#] & -7.3$^{+0.2}_{-0.3}$ & -7.3$^{+0.1}_{-0.2}$\\
         log VMR K [\#] &  -10.0$^{+0.7}_{-0.7}$ & -8.0$^{+0.1}_{-0.1}$\\
         Gem$_b$ [\#] & -35.89$\pm$0.46 [-34.82] & -34.242$\pm$0.004 [-34.82] \\
         MIRI1$_b$ [\#] &  -35.39$\pm$0.03 [-35.36] & -35.30$\pm$0.03 [-35.36] \\
         MIRI2$_b$ [\#] &  -36.11$\pm$0.03 [-35.98] & -36.12$\pm$0.03 [-35.98] \\
         MIRI3$_b$ [\#] &  -36.74$\pm$0.03 [-36.67] & -36.73$\pm$0.02 [-36.67] \\
         
    \end{tabular}
    \end{center}
    Notes: Values in square brackets are maximum likelihood values adopted in the full resolution retrievals. We do not report values on Na as the parameter was not constrained and we excluded it in the volume mixing ratio and metallicity calculations.  
   
\end{table}

\end{appendix}

\end{document}